\documentclass{elsart}
\journal{Nuclear Physics B}
\usepackage{graphicx}
\usepackage{amsmath}

\newcommand{\la}{\label}
\newcommand{\be}{\begin{equation}}
\newcommand{\ee}{\end{equation}}
\newcommand{\bea}{\begin{eqnarray}}
\newcommand{\eea}{\end{eqnarray}}

\def\la{\label}
\def\be{\begin{equation}}
\def\beq{\begin{equation}}
\def\eeq{\end{equation}}
\def\ee{\end{equation}}
\def\bea{\begin{eqnarray}}
\def\eea{\end{eqnarray}}
\def\p{\partial}

\begin{document}

\begin{frontmatter}

\title{Normal random matrix ensemble as a growth problem}

\author[la]{R.~Teodorescu \thanksref{now}}, 
\ead{rteodore@uchicago.edu}
\author[lb]{E.~Bettelheim \thanksref{now1}},
\ead{eldadb@phys.huji.ac.il}
\author[lb]{O.~Agam},
\author[lc]{A.~Zabrodin}, 
\ead{zabrodin@itep.ru} and
\author[ld]{P.~Wiegmann} 
\ead{wiegmann@uchicago.edu}
\thanks[now]{Present address: Physics Department, Columbia University, New York.}
\thanks[now1]{Present address: James Frank Institute, University of Chicago, 5640 S. Ellis Ave. Chicago, IL 60637, USA.}
\address[la]{James Frank Institute, University of Chicago, 5640 S. Ellis Ave. Chicago, IL 
60637, USA.}
\address[lb]{Racah Institute of Physics, Hebrew University, Givat Ram, Jerusalem, Israel 
91904}
\address[lc]{Institute of Biochemical Physics, Kosygina str. 4, 117334 Moscow, Russia also at ITEP, Bol. Cheremushkinskaya str. 25,
117259 Moscow, Russia}
\address[ld]{James Frank  Institute, Enrico Fermi Institute, University of Chicago, 5640 S. Ellis Ave. Chicago, IL and the Landau
Institute, Moscow, Russia}

\begin{abstract}
In general or normal random matrix ensembles,
the support of eigenvalues of large size matrices is a planar domain
(or several domains) with a sharp boundary.
This domain evolves under a  change
of parameters of the potential and of the size of matrices.
The boundary of the support of eigenvalues is a real section
of a complex curve. Algebro-geometrical properties
of this curve encode  physical properties of random matrix ensembles.
This curve can be treated as a limit of a spectral curve
which is canonically defined for models of finite matrices.
We interpret the evolution of the eigenvalue distribution
as a growth problem, and describe the growth in terms of evolution
of the spectral curve. We discuss algebro-geometrical
properties of the spectral curve and describe the wave functions
(normalized characteristic polynomials) in terms
of differentials on the curve. General formulae and emergence
of the spectral curve are illustrated by three meaningful examples.
\end{abstract}

\begin{keyword}
Integrable Systems \sep Random Matrix Theory \sep Stochastic Growth 

\PACS 02.50.Ey \sep 02.30.-f
\end{keyword} 

\end{frontmatter}

\section{Introduction and Preliminaries}
       \label{sec:intro}
Recently, random matrix theory has found
new applications in growth problems, where the evolution
of the interface separating domains of different nature is a subject
of interest. In some realizations, the growing domain is an aggregate of randomly
deposited subunits.

In general random matrix ensembles, complex  eigenvalues
usually occupy planar
domains in the complex plane. Their boundaries
become sharp as the size of matrices, $N$, goes to infinity.
It appears that, for an important class of growth models, the aggregates evolve
similarly to the support of eigenvalues of general matrix ensembles.
In Refrs. \cite{mwz,ABWZ02,mkwz}, one of the most interesting classes
of growth problems -- Laplacian growth -- has been  linked to the
evolution of normal random matrices.

The interpretation of random matrix
theory in terms of an aggregation process seems to be a
productive approach
in a number of different applications. Among them are the growth problems
mentioned above, the
semiclassical behavior of electronic droplets in the Quantum Hall
regime \cite{ABWZ02}, and ${\mathcal N}=1$
supersymmetric Yang-Mills theory \cite{DV03}.
They reveal a relatively new, geometrical aspect of
random matrix theory. From the mathematical point of view,
the connection of random matrix ensembles
with isomonodromic deformations of differential
equations \cite{BEH,Kapaev,gravity,Its} seems to be especially
interesting for these applications.

It has been emphasized in
recent papers
\cite{DV03} and \cite{KM03} that the boundary of the domains of
eigenvalues is to be thought of as a real section
of a complex algebraic curve (a Riemann surface). This curve
appears to be a fundamental object
of random matrix theory.
It encodes the most interesting physical
properties of matrix ensembles (see, e.g., \cite{DV03,KM03,Chekhov,David}
for the complex curves in one- and two-matrix models).

In the Hermitian or unitary ensembles, the support
of eigenvalues is a set of line intervals.
Their boundaries are just points.
In  general matrix ensembles, eigenvalues are complex numbers, and their
  support
is a planar domain. The boundaries are planar curves, evolving
in a complicated and unstable manner.
The notion of  complex curve
naturally links the planar geometry of the boundaries to
the algebro-geometric properties of the matrix ensemble.

   In this paper we present
general aspects of the normal matrix ensemble from this standpoint.
One may canonically associate a complex curve
to the matrix ensemble, not only in the large
$N$ limit, but for any finite $N$ as well.
It is the spectral curve of
the operator which generates recursion relations
(the Lax operator),
projected to a certain finite-dimensional subspace.
In the large $N$ limit (a semiclassical limit),
this curve becomes the complex curve which describes
the support of eigenvalues.

We introduce the spectral curve for normal matrix ensembles, and
describe the evolution of the curve
with respect to parameters of the statistical weight of the
ensemble {\it (deformation parameters)}
and the size of matrices (a parameter of
{\it growth}).
In a forthcoming paper, we will prove that the semiclassical limit of the evolution
(or deformation) equations
is the universal Whitham hierarchy associated with the complex curve.
The semiclassical wave function can be expressed
through differentials on the curve. We also notice that the
Whitham hierarchy is identical to the set of equations
which describe Laplacian growth processes --
unstable dynamics of an interface
between two immiscible phases.

Three meaningful examples illustrate the general formulae.

\subsection{Normal matrix ensemble}

A matrix $M$ is called normal if it commutes with its
Hermitian conjugate: $[M, M^{\dag}]=0$, so that both $M$ and $M^{\dag}$
can be diagonalized simultaneously. Eigenvalues of normal matrices
are complex.
The statistical weight of the normal matrix ensemble is given through a
general potential $W(M,M^{\dag})$ \cite{Zaboronsky}:
\begin{equation}
\label{ZN}
e^{\frac{1}{\hbar}{\rm tr} \, W(M, M^{\dag})} d\mu (M).
\end{equation}
Here $\hbar$ is a parameter, and the measure of integration
over normal matrices
is induced by the flat metric
on the space
of all complex matrices $d_C M$, where
$d_C M = \prod_{ij}d\, {\mathcal R}e \, M_{ij} d\, {\mathcal I}m \, M_{ij}$.
Using the standard procedure,
(see, e.g., \cite{Mehta91}) one passes to the joint
probability distribution
of eigenvalues of normal matrices $z_1,\dots,z_N$, where $N$ is
size of the matrix:
\begin{equation}
\label{mean}
\frac{1}{N! \tau_N}|\Delta_N (z)|^2 \,\prod_{j=1}^N
e^{\frac{1}{\hbar}W(z_j,\bar z_j)} d^2 z_j
\end{equation}
Here
$d^2 z_j \equiv dx_j \, dy_j$ for $z_j =x_j +iy_j$,
$\Delta_N(z)=\det (z_{j}^{i-1})_{1\leq i,j\leq N}=
\prod_{i>j}^{N}(z_i -z_j)$
is the Vandermonde determinant, and
\begin{equation} \label{tau}
\tau_N = \frac{1}{N!}\int
|\Delta_N (z)|^2 \,\prod_{j=1}^{N}e^{\frac{1}{\hbar}
W(z_j,\bar z_j)} d^2z_j
\end{equation}
is the normalization factor. This is the partition function
of the matrix model (a $\tau$-function).

A particularly important  special  case arises
if the potential $W$ has the form
\begin{equation}
\label{potential}
W=-|z|^2+V(z)+\overline{V(z)},
\end{equation}
where $V(z)$ is a holomorphic function in a domain which
includes the support of eigenvalues (see also a comment in the end of
Sec.~\ref{F1}
about a proper definition of the ensemble with this potential).
In this case, a normal matrix ensemble gives the same distribution as
       a general complex matrix ensemble.
A general complex matrix can be decomposed as $M=U(Z+R)U^\dagger$,
where $U$ and  $Z$
are  unitary and diagonal matrices, respectively, and $R$
is an upper triangular matrix. The distribution (\ref{mean})
holds for the elements of the diagonal matrix $Z$
which are eigenvalues of $M$ (Ref. \cite{Mehta91}).
Here we mostly focus on
the special potential (\ref{potential}),  and also assume
that the field
\be\la{a}
A(z)=\p_z V(z)
\ee
(a ``vector potential'', see Sec. \ref{QH})
is a globally defined meromorphic function.

\subsection{Droplets of eigenvalues}
\begin{figure} \begin{center} 
\includegraphics*[width=5cm]{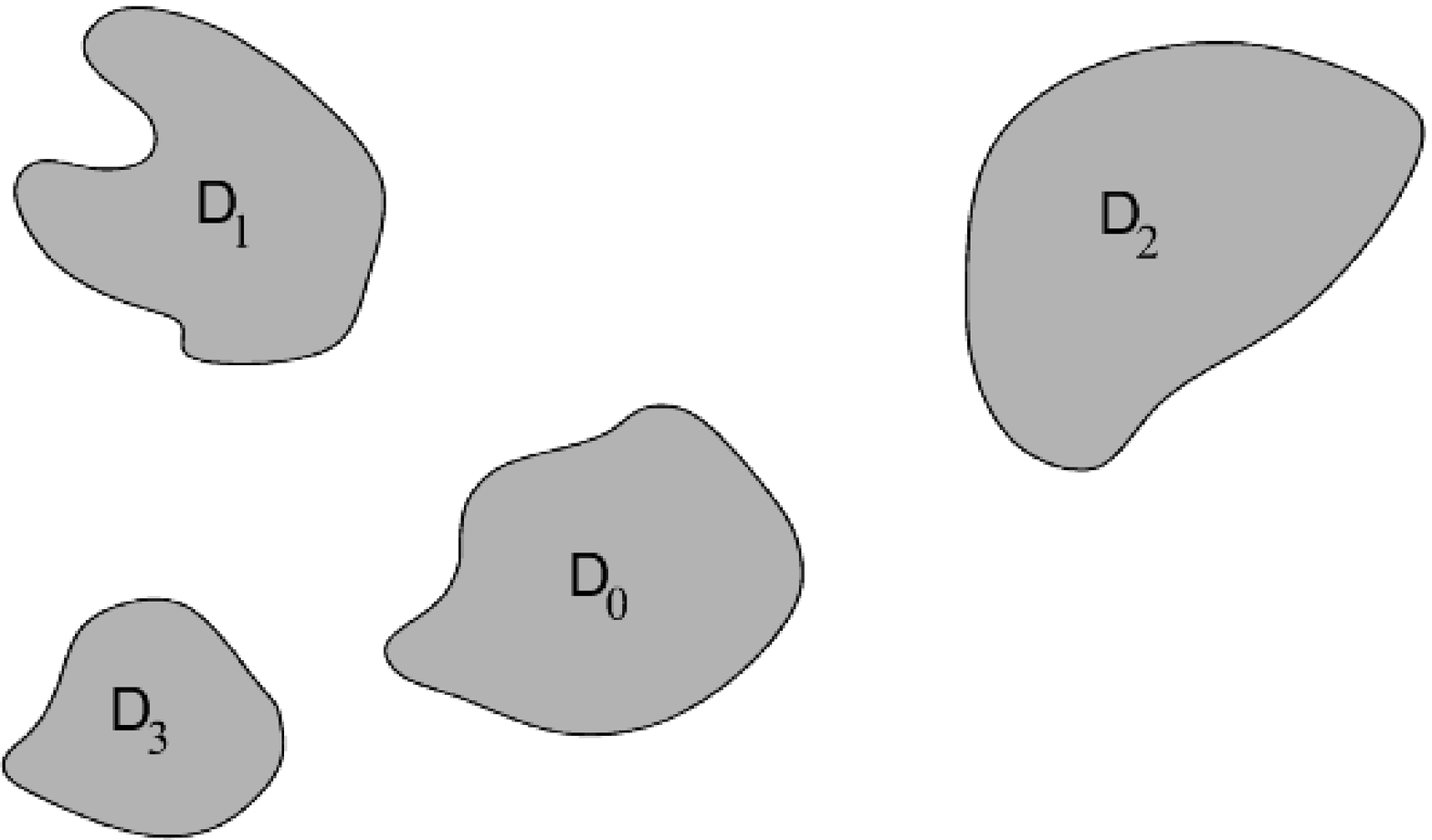}
\caption{A support of eigenvalues consisting of four disconnected components.}
\label{droplets}
\includegraphics*[width=5cm]{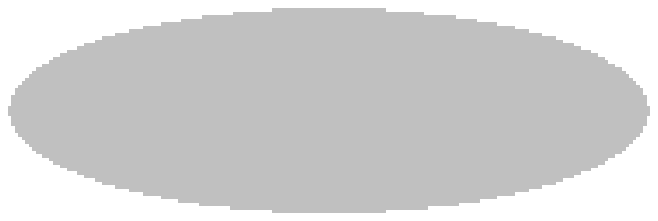}
\caption{The distribution of eigenvalues for the Gaussian potential. The droplet is an ellipse with quadrupole moment $2|t_2|$ and area $\pi\hbar N$.}
\label{ellipse}
\end{center} \end{figure}

     In a proper large $N$ limit ($\hbar\to 0$,
$N\hbar$ fixed), the eigenvalues of matrices from the
ensemble densely occupy  a connected domain $D$
       in the complex plane, or, in general, several
disconnected domains.
This set (called the support of eigenvalues)
has sharp edges
(Fig.~\ref{droplets}). We refer to the connected components
$D_\alpha$ of the domain $D$ as {\it droplets}.

\begin{figure} \begin{center}
      \includegraphics*[width=5cm]{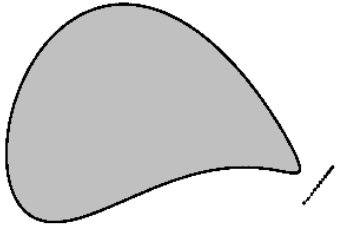}
\caption{The distribution of eigenvalues for the potential $V(z) = - \alpha \log (1- z/\beta) - \gamma z $.
If the area of the droplet is $\pi\hbar N$, the complement of the droplet is an algebraic domain which can be conformally
mapped to the exterior of the unit disk by the Joukowsky map (\ref{222}). In this case, the cut of the Schwarz function
  shown in the figure (corresponding  to a virtual droplet located on the unphysical sheet)   shrinks to a double point. }
\label{Joukowsky}
\end{center} \end{figure}

\begin{figure} \begin{center}
      \includegraphics*[width=5cm]{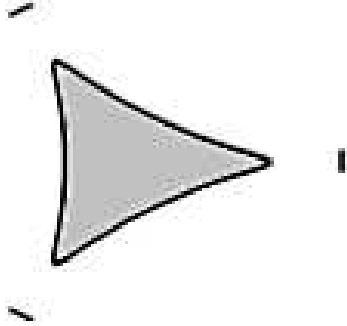}
\caption{The distribution of eigenvalues for the potential
   $V(z) = t_3z^3$.  Cuts of the Schwarz function are shown. The cuts
correspond to virtual droplets located on unphysical sheets.
If the area of the droplet is $\pi\hbar N$, the cuts shrink to double points.
The boundary contour is a hypotrochoid.}
\label{hypotrochoid}
\end{center} \end{figure}

In the case of algebraic domains (the definition follows)
the eigenvalues are distributed with the density
$\rho=-\frac{1}{4 \pi}\Delta W$,
where $\Delta =4\p_z \p_{\bar z}$ is the 2-D Laplace operator
\cite{WZ03}.
For the potential (\ref{potential}) the density is uniform.
The shape of the support
of eigenvalues is a more involved subject. We discuss it below.
For example, if the
potential is Gaussian \cite{Ginibre65},
\be\la{e}
A(z)=2t_2 z,
\ee
the domain is an ellipse (see Fig.~\ref{ellipse}).
If $A$ has one simple pole,
\be\la{J1}
A(z)=-\frac{\alpha}{z-\beta}-\gamma
\ee
the droplet (under certain conditions discussed below) has the profile
of an aircraft wing given by the Joukowsky map
(Fig.~\ref{Joukowsky}).
If $A$ has one double pole
(say, at infinity),
\be\la{hyp}
A(z)=3t_3 z^3,
\ee
the droplet is a hypotrochoid
(Fig.~\ref{hypotrochoid}).
These are the three examples that we will discuss below in detail.
If $A$ has two or more simple poles, there may be
more than one droplet.

\begin{figure} \begin{center}
      \includegraphics*[width=5cm]{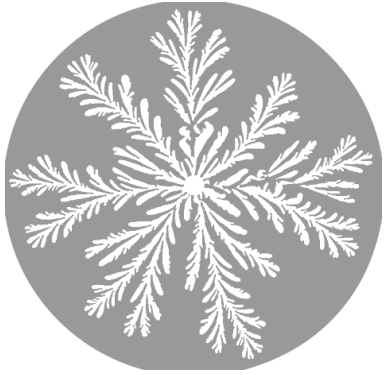}
      \caption{\label{sw}
A grown fingering pattern observed
in a radial Hele-Shaw cell.
Air is inserted under pressure to the cell filled by
  silicon oil \cite{Sw}.}
\end{center} \end{figure}

If $A(z)$ is a more complicated function,
the domain of eigenvalues develops
an unstable fingering pattern,
similar to the one in Fig.~\ref{sw}.

Boundary components of the droplets form a real section of
a complex curve.

\subsection{Semiclassical complex curve}
Jumping ahead, we describe the  construction of the semiclassical
   complex curve which emerges in the
normal matrix ensemble. Let us
represent the boundary of the domain
as a real  curve
$F(x,\,y)=0$. If the vector potential $A(z)$ is a meromorphic
function (we always assume that this is the case), the function
$F$ can be chosen to be an irreducible polynomial.
Then we rewrite it
in holomorphic coordinates as
\be\label{F}
F\left (\frac{z+\bar z}{2},\frac{z-\bar z}{2i}\right )=
f(z,\bar z)
\ee
and treat $z$ and $\bar z$ as independent complex coordinates
$z$, $\tilde z$.
The equation
$f(z,\tilde z)=0$ defines a complex curve.
This curve is a finite-sheet covering of the $z$-plane.
The single-valued function $\tilde z(z)$
on the curve is a multivalued function on
the $z$-plane. Making cuts, one can fix single-valued
branches of this function.

The boundary of the domain is a section of the curve by the plane where
$\tilde z$ is complex conjugate of $z$.
It belongs to a particular
sheet (we call it the {\it physical sheet}) of the covering.
It appears that physical properties of the ensemble are
determined not only by the physical sheet,
but rather by the entire algebraic
covering, including all the sheets other than physical. The Riemann
surface for 
Example (\ref{J1}) is presented in Fig.~\ref{torus}.

\subsection{Growth and deformation parameters. Evolution of the curve}
Similarly to the Hermitian matrix ensemble
\cite{DV03,KM03,Chekhov,David},
the complex  curve of the normal matrix ensemble
   is characterized by the potential $W$, or by
the ``vector potential"
$A(z)$, and  by a set of $g+1$
integers $\nu_\alpha$ (not necessarily positive), where $g$ is
genus of the curve.
The integers are subject to the constraint
$\sum_{\alpha=0}^g\nu_\alpha=N$.
We will discuss the meaning of these numbers later on.
   For now, we only mention that if they are all positive, then they
   are proportional to the areas of the droplets
of uniformly distributed eigenvalues. Every droplet contains
$\nu_\alpha$ eigenvalues, so a quantum of area is
$\pi\hbar$ per particle.

As one varies the potential
and the filling factors $\nu_\alpha$,
the curve and the interface bounding the droplets evolve.
Parameters of the potential
(for example, poles and residues of the meromorphic function (\ref{a}))
and filling numbers are
{\it deformation parameters} and {\it parameters of growth}.
They are coordinates
in the moduli space of the complex curves.

An infinitesimal variation of the potential
generates  correlation functions of the  ensemble
(Ref. \cite{WZ03}). For example,
irreducible correlation functions
of resolvents (holomorphic currents)
$J (z)=\mbox{tr}\, \frac{\hbar}{z-M}=\sum_i\frac{\hbar}{z-z_i}$
are generated by variations of $A(z)$.
In Ref. \cite{WZ03} we showed that the correlation functions
are expressed through algebro-geometrical properties of the boundary of
the domain.

During the evolution, the genus of the
complex curve may change. This
corresponds to  a coalescence of droplets or a droplet breakup, or
even more complicated
degenerations, when a droplet shrinks to a point
and simultaneously merges with another droplet.
Close to degeneracy points (also called critical points), matrix
ensembles have universal scaling behavior.
We will present a study of this problem in a future paper.

\begin{figure} \begin{center}
   \includegraphics*[width=5cm]{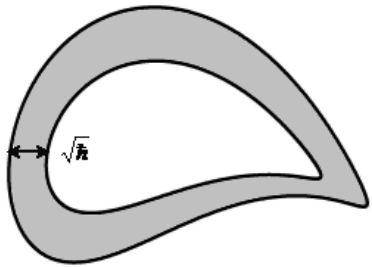}
   \caption{\label{semi}The area of a droplet grows with the rate  $\pi \hbar$
per eigenvalue. A new eigenvalue aggregates at the boundary of the
droplet. The shaded area
depicts the support of the semiclassical  wave function.}
\end{center} \end{figure}

\begin{figure} \begin{center}
   \includegraphics*[width=5cm]{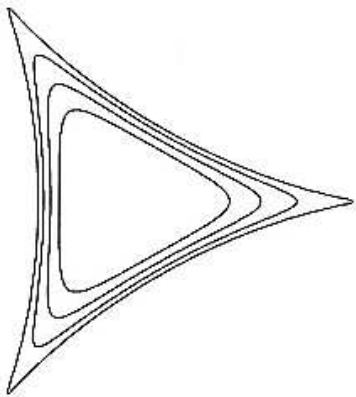}
      \caption{\label{evolutionhyp}Hypotrochoid grows
until it reaches a critical point.}
\end{center} \end{figure}

A particularly interesting process is {\it growth}. In this case, one
changes the
total number of eigenvalues $N$
     and keeps the potential fixed. Below we refer to the normalized
total number of particles
$$t=\hbar N$$
as {\it time of growth}. In the algebraic case, the time $t$ is
the normalized (modulo $\pi$) area of the droplets. While $N$
  increases, the new eigenvalues aggregate at the boundary of the
existing droplets,
so that the domain of eigenvalues  (an aggregate) grows
(Fig.~\ref{semi}).
Growth of a hypotrochoid is depicted in Fig.~\ref{evolutionhyp}.

There are several important physical problems where this kind of evolution
occurs. We list them in historical order.

The first one is the celebrated  Hele-Shaw problem  -- the most studied
example of  Laplacian growth.
The Hele-Shaw problem  describes the non-equilibrium
dynamics  of a planar interface between two immiscible fluids
confined to a thin 2-D cell.
The second one is the matrix
model description of  2-D quantum gravity.
The third one is the evolution of energy levels in mesoscopic systems.
A similar  problem also emerges in isomonodromic deformation of
differential equations.

The last three problems and their connection with the matrix ensemble
are well reflected
in the literature (see  e.g., \cite{gravity,Its,mesoscopic} for reviews).
The connection of matrix ensembles to the  Hele-Shaw
problem is relatively new.
It nicely
illustrates the growth and gives
a transparent hydrodynamic interpretation to abstract
objects of algebraic geometry of curves and complex analysis on
Riemann surfaces.
We find that a brief description of the Hele-Shaw problem in this
paper is in order (see Sec. \ref{Appendix_A}).

\subsection{Normal matrix ensemble
and quantum Hall effect}\la{QH}

A useful interpretation of the  Coulomb gas distribution (\ref{mean}) is
a coherent state of  relativistic electrons in the Quantum Hall regime
\cite{ABWZ02}. In this case, the electrons are
situated in the plane in a strong, not necessarily
uniform, magnetic field
$B(z)=-\frac{1}{2} \Delta W$, and fully occupy the lowest
energy level. The exact $N$-particle wave function, defined up to a phase, is
\begin{equation}\label{psi}
\Psi_N(z_1,\dots,z_N)=
\frac{1}{\sqrt{N! \tau_N}}\Delta_N (z) \,e^{\sum_{j=1}^{N}
\frac{1}{2\hbar}W(z_j, \bar z_j)}.
\end{equation}
The joint probability distribution (\ref{mean})
is then equal to $|\Psi(z_1,\dots,z_N)|^2$.
It is the  probability to find electrons
at the points $z_i$. In the semiclassical
regime, the wave function (\ref{psi})
describes an incompressible electronic droplet with sharp edges.
The function $A(z)$ (\ref{a}) can be thought of as
a holomorphic component of the vector potential
generated by magnetic impurities located far away from the
electronic droplet.

In this language, the growth problem translates into the evolution of
a semiclassical
electronic droplet under a change of magnetic
field or chemical potential.

\subsection{Orthogonal polynomials as a measure of growth}\la{F1}
Let the number of eigenvalues (particles)
increase while the potential  stays fixed. If the
support of eigenvalues is
simply-connected, its  area
grows as $\hbar N$.  One can describe the
evolution of the  domain through the density of particles
\begin{equation}\la{51}
\rho_N(z)=N \int|\Psi_{N}(z,z_1,z_2,\dots,z_{N-1})|^2
d^2z_1\dots d^2 z_{N-1},
\end{equation}
where $\Psi_N$ is given by (\ref{psi}).

We introduce a set of orthonormal one-particle functions
on the complex plane as matrix elements of transitions
between $N$ and $(N+1)$-particle states:
\begin{equation}\la{5}
\frac{\psi_N(z)}{\sqrt{N+1}}= \int\Psi_{N+1} (z,z_1,z_2,\dots,z_N)
\overline{\Psi_{N}(z_1,z_2,\dots,z_N)} d^2z_1\dots d^2 z_N
\end{equation}
Then the rate of the density change is
\begin{equation}\la{1}
\rho_{N+1}(z)-\rho_{N}(z)=|\psi_N(z)|^2.
\end{equation}
The proof of this formula
is based on the representation of the $\psi_n$ through
holomorphic biorthogonal polynomials $P_n(z)$. Up to a phase
\begin{equation}\label{O3}
\psi_n (z)=
e^{\frac{1}{2\hbar}W(z, \bar z )}P_n (z),\quad
\quad P_n (z)=\sqrt{\frac{\tau_n}{\tau_{n+1}}} z^n +\ldots
\end{equation}
The polynomials $P_n(z)$ are
biorthogonal on the complex plane
with the weight $e^{W/\hbar}$:
\begin{equation}\la{29}
\int e^{W/\hbar}P_n(z)\overline{P_m(z)}d^2 z=
\delta_{mn}
.
\end{equation}
The proof of these formulae
is standard in the theory of orthogonal
polynomials. Extension to the biorthogonal case adds no
difficulties.

A physical interpretation of the
wave function $\psi_n$ is clear in the QHE
setup. The eigenvalues $z_i$
are the positions of electrons in a strong magnetic field.
Then $|\psi_N(z)|^2$ is
  a probability of adding an additional electron to
an aggregate  of $N$ electrons at the point $z$. Since the droplet is
incompressible, it
is only  possible to add the particle to the boundary of the droplet.
We will see that the  wave function
in a semiclassical limit is indeed localized on the boundary.

We note that, with
   the choice of potential (\ref{potential}), the integral
representation (\ref{29}) has only a formal meaning,
since the integral diverges
unless the potential is Gaussian.
A proper definition of the  wave functions goes through
recursive relations (\ref{L1}, \ref{M}) which follow from the integral
representation.
The same comment applies to the $\tau$-function (\ref{tau}).
The wave function is not  normalized everywhere
in the complex plane. It may diverge at the poles of the
vector potential field.

Below we describe the evolution and
      semiclassical behavior
of the wave function in the region of the complex plane close to the
boundary of physical droplets.

\section{Equations for the wave functions and the
spectral curve}\la{1A}
In this section we specify the potential to be
of the form (\ref{potential}).
It is convenient to modify the
exponential factor of the wave function.
Namely, we define
\be \la{chin}
\psi_n (z)=
e^{-\frac{|z|^2}{2\hbar}+\frac{1}{\hbar}V(z)}P_n (z),\quad\mbox{and}\quad\chi_n (z)=
e^{\frac{1}{\hbar}V(z)}P_n (z),
\ee
where the holomorphic
functions $\chi_n(z)$
are orthonormal in the complex plane
with the weight
$e^{-|z|^2/\hbar}$.
Like traditional orthogonal polynomials, the biorthogonal polynomials
$P_n$ (and the corresponding wave functions)
obey a set of differential equations with
respect to the argument $z$, and recurrence relations with respect to
the degree $n$. Similar equations
for  two-matrix models are discussed in  numerous papers
(see, e.g., \cite{Aratyn}).

We introduce the $L$-operator (the Lax operator)
as multiplication by $z$ in the
basis $\chi_n$:
\begin{equation}\la{L1}
L_{nm}\chi_m(z)=z\chi_n(z)
\end{equation}
(summation over repeated indices is implied).
Obviously, $L$ is a lower triangular matrix with one
adjacent upper diagonal, $L_{nm}=0$ as $m>n+1$.
Similarly, the differentiation
$\p_z$ is represented by an
upper triangular matrix with one adjacent lower diagonal.
Integrating by parts the matrix elements of the $\p_z$, one finds:
\begin{equation}\la{M}
(L^{\dag})_{nm}\chi_m =
\hbar\p_z\chi_n,
\end{equation}
where $L^{\dag}$
is the Hermitian conjugate operator.

The matrix elements of $L^{\dag}$ are
$(L^{\dag})_{nm}=\bar L_{mn}=A(L_{nm})+
\int e^{\frac{1}{\hbar}W}\bar P_m(\bar z)\p_z P_n(z) d^2z$, where
the last term is a lower triangular matrix.  The latter can be written
       through negative powers of the Lax operator.
Writing $\p_z\log P_n(z)=\frac{n}{z}+\sum_{k>1}v_k(n)z^{-k}$,
one represents $L^{\dag}$ in the form
\begin{equation}\la{M1}
L^{\dag}=A(L)+(\hbar n) L^{-1}+\sum_{k>1}v^{(k)}L^{-k},
\end{equation}
where $v^{(k)}$ and $(\hbar n)$ are
diagonal matrices with elements $v_n^{(k)}$
and $(\hbar n)$.
The coefficients $v_{n}^{(k)}$ are determined by the condition that
lower triangular matrix elements of $A(L_{nm})$ are cancelled.

In order to emphasize the structure of the operator $L$, we
write it in the basis of the
shift operator \footnote{The shift operator
$\hat w$  has no
inverse. Below $\hat w^{-1}$ is understood as a shift to the left
defined as  $\hat w^{-1}\hat w=1$.
Same is applied to the operator $L^{-1}$. To avoid a possible
confusion, we emphasize that although $\chi_n$
is a right-hand eigenvector of $L$, it is not a right-hand
eigenvector of $L^{-1}$.}
$\hat w$ such that $\hat w f_n =f_{n+1}\hat w$ for any
sequence $f_n$. Acting on the wave function, we have:
$$\hat w
\chi_n=\chi_{n+1}.$$
In the $n$-representation, the operators $L$, $L^{\dag}$
acquire the form
\begin{equation}\la{M11}
L=r_n \hat w+\sum_{k\geq 0} u_{n}^{(k)} \hat w^{-k},\quad
L^{\dag} = \hat w^{-1} r_n+
\sum_{k\geq 0} \hat w^{k} \bar u_{n}^{(k)}.
\end{equation}
Clearly, acting on $\chi_n$, we have the commutation
relation (``the string equation")
\begin{equation}\la{string}
[L, \, L^{\dag}]=\hbar.
\end{equation}
This is the compatibility condition of Eqs. (\ref{L1}) and (\ref{M}).

Equations (\ref{M11}) and (\ref{string})
completely determine the coefficients
$v_n^{(k)}$, $r_n$ and $u_n^{(k)}$. The first one connects the coefficients
to the parameters of the potential.
The second equation is used to determine how the coefficients
       $v_n^{(k)}$, $r_n$ and $u_n^{(k)}$ evolve with $n$. In particular,
the diagonal part of it reads
\be\la{Area}
n\hbar=r_n^2-\sum_{k\geq 1}\sum_{p=1}^k|u_{n+p}^{(k)}|^2.
\ee
Moreover, we note that all the coefficients can be expressed
through the $\tau$-function (\ref{tau}) and its derivatives
with respect to parameters of the potential.
This representation is particularly simple for $r_n$:
$
r_n^2 =\tau_n \tau_{n+1}^{-2}\tau_{n+2}
$.

\subsection{Finite dimensional reductions}
If the vector potential $A(z)$ is a rational function,
the coefficients $u_{n}^{(k)}$ are not all independent.
The number of independent coefficients
equals the number of independent parameters of the potential.
For example, if the holomorphic part
of the potential, $V(z)$, is a polynomial of degree $d$,
the series
(\ref{M11}) are truncated at $k= d-1$.

In this  case the semi-infinite system of linear equations (\ref{M})
and the recurrence relations (\ref{L1})
can be cast in the form of a set of finite dimensional equations whose
coefficients are rational functions of  $z$, one
system for every $n>0$.
The system of differential equations generalizes the
Cristoffel-Daurboux second order
differential equation
valid for orthogonal polynomials. This fact has been observed in
recent papers \cite{BEHdual,Eynard03}
for biorthogonal polynomials emerging in the
Hermitian two-matrix model
with a polynomial potential. It is
applicable to our case (holomorphic biorthogonal
polynomials) as well.

In a more general case,
when $A(z)$ is a general rational function with $d-1$ poles
(counting multiplicities), the series (\ref{M11})
is not truncated. However, $L$ can be represented
as a ``ratio",
\beq\label{LKK}
L=K_{1}^{-1}K_{2}=M_2 M_{1}^{-1},
\eeq
where the operators $K_{1,2}$, $M_{1,2}$ are
polynomials in $\hat w$:
\beq\label{KK}
K_1 =\hat w^{d-1}+\sum_{j=0}^{d-2}A_{n}^{(j)} \hat w^j\,,
\quad
K_2 =r_{n\! +\! d\! -\! 1}\hat w^{d}+\sum_{j=0}^{d-1}B_{n}^{(j)} \hat w^j
\eeq
\beq\label{MM}
M_1 =\hat w^{d-1}+\sum_{j=0}^{d-2}C_{n}^{(j)} \hat w^j\,,
\quad
M_2 =r_{n}\hat w^{d}+\sum_{j=0}^{d-1}D_{n}^{(j)} \hat w^j
\eeq
These operators obey the
relation
\beq\label{KMKM}
K_1 M_2 =K_2 M_1.
\eeq
It can be proven that
the pair of operators $M_{1,2}$ is uniquely determined
by $K_{1,2}$ and vice versa. We note that the
reduction (\ref{LKK}) is a difference analog of the
``rational" reductions of the Kadomtsev-Petviashvili
integrable hierarchy considered in \cite{Krichev-red}.

The linear problems (\ref{L1}), (\ref{M}) acquire the form
\beq\label{L1a}
(K_2\chi )_n =z \, (K_1 \chi )_n\,, \quad
(M_{2}^{\dag} \chi )_n =\hbar \p_z (M_{1}^{\dag}\chi ) _n.
\eeq
These equations are of {\it finite order}
(namely, of order $d$), i.e., they connect values of $\chi_n$
on $d+1$ subsequent sites of the lattice.

The semi-infinite set $\{\chi_0,\chi_1,\dots\}$ is
then a ``bundle" of
$d$-dimensional vectors
$${\underline\chi}(n)= (\chi_n,\chi_{n+1},\dots,
\chi_{n+d-1})^{{\rm t}}$$
(the index ${\rm t}$
means transposition, so
${\underline\chi}$ is a column vector).
The dimension of the vector is the number of poles
of $A(z)$ plus one.
Each vector obeys a closed $d$-dimensional linear
differential equation
\begin{equation}
\la{M2}
\hbar\p_z{\underline\chi}(n)={\mathcal L}_n (z){\underline\chi}(n),
\end{equation}
where the $d\times d$ matrix
${\mathcal L}_n$ is a ``projection" of the operator $L^{\dag}$
onto the $n$-th $d$-dimensional space. Matrix elements of the
${\mathcal L}_n$ are rational functions of
$z$ having the same poles as $A(z)$ and also a pole at
the point $\overline{A(\infty )}$. (If $A(z)$ is a polynomial,
all these poles accumulate to a multiple pole at infinity).

We briefly describe the procedure of
constructing the finite dimensional matrix
differential equation. We use the first
linear problem in (\ref{L1a})
to represent the shift operator
as a $d\times d$ matrix ${\mathcal W}_n (z)$ with
$z$-dependent coefficients:
\be\la{W}
{\mathcal W}_n(z){\underline\chi}(n)={\underline\chi}(n \! +\! 1).
\ee
This is nothing else than rewriting the scalar linear problem
in the matrix form.
Then the matrix ${\mathcal W}_n (z)$ is to be substituted into the
second equation of (\ref{L1a}) to
determine ${\mathcal L}_n(z)$ (examples follow).
The entries of ${\mathcal W}_n(z)$ and ${\mathcal L}_n (z)$
obey the Schlesinger
equation, which follows from compatibility of
(\ref{M2}) and (\ref{W}):
\be\la{WW}
\hbar\p_z {\mathcal W}_n=
{\mathcal L}_{n+1} {\mathcal W}_n
-{\mathcal W}_n {\mathcal L}_n.
\ee

This procedure has been realized explicitly for
polynomial potentials in
recent papers \cite{BEHdual,Eynard03}.
We will work it out in detail for our three examples:
${\underline\chi}(n)=(\chi_n,\,\chi_{n+1})^{{\rm t}}$ for the ellipse
(\ref{e}) and
the aircraft wing (\ref{J1}) and
${\underline\chi}(n)=(\chi_n,\,\chi_{n+1},\chi_{n+2})^{{\rm t}}$
for the hypotrochoid (\ref{hyp}).

\subsection{Spectral curve}\la{Curve1}

According to the general theory of
linear differential equations, the
semiclassical (WKB) asymptotics of solutions to
Eq. (\ref{M2}), as $\hbar \to 0$,
is found by solving the eigenvalue problem for
the matrix ${\mathcal L}_n (z)$ (see, e.g., \cite{Wasow}) .
More precisely, the basic object of the WKB approach is the
spectral curve \cite{curve} of the matrix ${\mathcal L}_n$, which is
defined, for every integer $n>0$, by the secular equation
$\det ({\mathcal L}_n (z)-\tilde z) = 0$
(here $\tilde z$ means $\tilde z \cdot {\bf 1}$, where
${\bf 1}$ is the unit $d\times d$ matrix).
It is clear that the left hand side of the secular
equation is a polynomial in $\tilde z$ of degree $d$.
We define the spectral curve by an equivalent equation
\begin{equation}\la{qc}
f_n (z,\tilde z)=a(z)\det ({\mathcal L}_n (z)-\tilde z) = 0,
\end{equation}
where the factor $a(z)$ is added
to make $f_n(z,\tilde z)$ a polynomial in $z$ as well.
The factor $a(z)$ then has zeros at the points where
poles of the matrix function ${\mathcal L}(z)$ are located.
It does not depend on $n$.
We will soon see that the degree of
the polynomial $a(z)$ is equal to $d$.
Assume that all poles of $A(z)$ are simple, then zeros of
the $a(z)$ are
just the $d-1$ poles of $A(z)$ and another simple zero
at the point $\overline{A(\infty)}$. Therefore,  we conclude that
the matrix ${\mathcal L}_n (z)$ is rather special.
For a general $d\times d$ matrix function with the same
$d$ poles, the factor $a(z)$ would be of degree $d^2$.

Note that the matrix ${\mathcal L}_n (z) -\bar z$ enters
the differential equation
\begin{equation}\la{M2'}
\hbar\p_{z}|{\underline\psi}(n)|^2=\bar {\underline\psi}(n)
({\mathcal L}_n (z)- \bar z)
{\underline\psi}(n)
\end{equation}
for the squared amplitude
$|{\underline \psi}(n) |^2={\underline \psi}^{\dag}(n)
{\underline \psi}(n) =
e^{-\frac{|z|^2}{\hbar}} |{\underline \chi}(n)|^2$
of the vectors
${\underline \psi}(n)$ built from the orthonormal
wave functions (\ref{O3}).

The equation of the curve can be interpreted as a
``resultant" of the non-commutative polynomials
$K_2 -zK_1$ and $M_{2}^{\dag}-\tilde z M_{1}^{\dag}$
(cf. \cite{BEHdual}). Indeed, the point $(z, \tilde z)$
belongs to the curve if and only if the linear system
\beq\label{linsys}
\left \{ \begin{array}{l}
(K_2 c)_k =z (K_1 c)_k \quad \quad n-d\leq k \leq n-1
\\ \\
(M_{2}^{\dag} c)_k =\tilde
z (M_{1}^{\dag} c)_k \quad \quad n\leq k \leq n+d-1
\end{array}\right.
\eeq
has non-trivial solutions. The system contains $2d$ equations
for $2d$ variables $c_{n-d}\, , \ldots , c_{n+d -1}$.
Vanishing of the
$2d \, \times \, 2d$ determinant yields the equation of
the spectral curve.
Below we use this method to find the equation of the curve
in the examples.
It appears to be much easier than the determination of the
matrix ${\mathcal L}_n(z)$.

The spectral curve (\ref{qc}) possesses an important property:
it admits an antiholomorphic involution.
In the coordinates $z, \tilde z$ the involution reads
$(z, \tilde z)\mapsto (\overline{\tilde z}, \bar z)$.
This simply means that the secular equation
$\det (\bar {\mathcal L}_n (\tilde z)-z) = 0$
for the matrix $\bar {\mathcal L}_n (\tilde z)\equiv
\overline{{\mathcal L}_n (\overline{\tilde z})}$
defines the same curve.
Therefore, the polynomial $f_n$
takes real values for $\tilde z =\bar z$:
\be\la{anti1}
f_n(z,\bar z)=\overline{f_n ( z, \bar z)}.
\ee
Points of the real section of the curve
($\tilde z =\bar z$) are
fixed points of the involution.

The curve (\ref{qc})
was discussed in recent papers \cite{BEHdual,Eynard03}
in the context of Hermitian two-matrix models with
polynomial potentials. The dual realizations of the curve
pointed out in \cite{BEHdual} correspond to the
antiholomorphic involution in our case.
The involution can be proven along the lines of these
works. The proof is rather technical and we omit
it, restricting ourselves to the
examples below. We simply note that the involution relies on the fact
that the squared modulus of the wave function is real.

\subsection{Schwarz function}

The polynomial $f_n(z, \bar z)$ can be
factorized in two ways:
\be\la{h1}f_n(z,\bar z)=a(z)(\bar z-S_n^{(1)}(z))
\dots (\bar z-S_n^{(d)}(z)),
\ee
where $S_n^{(i)}(z)$ are eigenvalues of the matrix
${\mathcal L}_n (z)$, or
\be\la{ah1}
f_n(z,\bar z)=\overline{a(z)}( z-\bar S_n^{(1)}(\bar z))\dots ( z-\bar
S_n^{(d)}(\bar z)),
\ee
where $\bar S_n^{(i)}(\bar z)$ are eigenvalues of the matrix
$\bar {\mathcal L}_n (\bar z)$.
One may understand them as different branches of a
multivalued function $S(z)$ (respectively, $\bar S(z)$)
on the plane (here we do not indicate the dependence on $n$,
for simplicity of the notation).
It then follows that
$S(z)$ and $\bar S(z)$
are mutually inverse functions:
\begin{equation}\la{anti11}
\bar S(S(z))=z.
\end{equation}

An algebraic function  with this property
is called {\it the Schwarz function}.
By the equation $f(z, S(z))=0$, it defines a complex curve
with an antiholomorphic involution. An upper bound
for genus of this curve is $g=(d-1)^2$, where $d$ is the
number of branches of the Schwarz function.
The real section of this curve is a set of all fixed
points of the involution. It consists of a number
of contours on the plane (and possibly a number of
isolated points, if the curve is not smooth).
The structure of this set is known to be
  complicated. Depending on
coefficients of the polynomial, the number of disconnected
contours in the real section may vary from $0$ to $g+1$.
If the contours divide the complex curve into two
disconnected ``halves", or sides (related by the involution), then
the curve can be realized as
the {\it Schottky double} \cite{Alhfors50,C,SS} of one of
these sides. Each side is a Riemann surface with a boundary.

We will discuss general properties of the Schwarz function and the
Schottky double in Sec. \ref{S}.

Let us come back to
equation (\ref{M2}). It has $d$ independent solutions.
They are functions on the spectral curve.
One of them is a physical solution
corresponding to biorthogonal polynomials.
The physical solution defines the ``physical sheet"
of the curve.

The Schwarz function on the physical sheet
is a particular root, say $S^{(1)}_n(z)$,   of the
polynomial $f_n(z, \tilde z)$ (see (\ref{h1})).
It follows from (\ref{M1}) that
this root  is
selected by the requirement that
it has the same poles and residues as the potential $A$.

A formal \footnote{This formal expression ignores
the Stokes phenomenon. It
is valid only around  boundaries of  physical droplets.}
semiclassical asymptote of the
solution, in the leading order in $\hbar$, is
\be\la{semi1}
\chi_n\sim e^{\frac{1}{\hbar}\int^z d\Omega^{(1)}_n}.
\ee
Here
$$d\Omega_n^{(1)}=S_n^{(1)}dz.$$
The differential $d\Omega^{(1)}$ is a physical branch of the generating
differential on the curve (see below).

The semiclassical asymptotics is
discussed in more details in Sec. \ref{Semi2}.
Here we notice  that the amplitude of the wave function
$\psi_n (z)$ ( not $\chi_n (z)$!) peaks at the solution
of the equation
\be\la{S1}
S^{(1)}(z)=\bar z,
\ee
where now  $z=x+iy$ and $\bar z=x-iy$ are complex conjugated
coordinates in the plane.
Solutions to this equation describe
those contours of a real section of the complex curve
which belong to the physical sheet.
It is a set of closed
planar curves, as shown in Fig.~\ref{droplets}.
These curves
   are boundaries of semiclassical droplets. Evolution and  growth  of
the droplets
translate  into evolution of the complex curve.

\subsection{Deformation parameters}
The essential  information about the spectral curve
(which, in what follows, we will call {\it quantum curve},
unless the semiclassical limit is considered)
is contained in two conditions:
\begin{itemize}
\item
the antiholomorphic involution, and
\item  poles and residues
of the Schwarz function on one selected sheet (the physical
sheet) are given by the vector potential field $A(z)$.
\end{itemize}
These requirements determine all but $g$ coefficients
of the polynomial $f_n(z,\bar z)$, where $g$ is genus of the curve.
The  remaining $g$ coefficients do not depend on the potential
(deformation parameters).

The Bohr-Sommerfield quantization condition for the semiclassical
wave function
(\ref{semi1}) requires that
the integral over all the cuts (or boundaries of the droplets) are integer
   multiples of $2\pi i$. Also, the integral around the point
at infinity, counting
the degree of the polynomial, must be $2\pi iN$. The integrals over
the cuts transform
into the integrals over $\bf a$-cycles of the curve. Later we
will interpret the $\bf a$-cycles
as boundaries of the droplets. Therefore, we have the
set of integers
\begin{equation}\la{18}
\nu_a=
\frac{1}{2\pi i \hbar}\oint_{{\bf a}_\alpha} S(z) dz,
\quad\sum_\alpha\nu_\alpha=N,\quad\alpha=0,\,1,\dots, g.
\end{equation}

   The $d$-dimensional linear differential equations  (\ref{M2})
give an interpretation of the potential data as deformation parameters.
Varying  the poles and residues of
  $A(z)$, one does not change the monodromy of solutions, given by $N$
and $\nu_\alpha$.
They are similar to deformation parameters of the theory of
isomonodromic deformations of the system of linear differential
equations  \cite{Miwa} (see also  Ref. \cite{Its} for  applications of the
   theory of isomonodromic deformations to Hermitian matrix models
   and orthogonal polynomials in one variable, and references therein).
A connection between isomonodromic deformations and
Whitham equations, which seems to be relevant
to our discussion, was pointed out in Ref. \cite{Takasaki}.

In the spirit of the theory of isomonodromic deformations of
linear differential equations, we  conjecture that a specific value of
$N$, a potential $V(z)$, and a set of
$g$ numbers (\ref{18}) form a complete data set which uniquely
determines the coefficients
$r_n,\,u^{(k)}_n$ of the recurrence equations, and the solution of the
differential equation (\ref{M2}). We do not know, however,
whether the requirement that the physical solutions are polynomials  gives a
restriction on
possible values of the integers $\nu_\alpha$. It is likely  that a  general
set of $\nu_\alpha$ (positive and negative)
corresponds to  more general solutions of (\ref{M2}) rather than polynomials.

We also conjecture that, for physical values
of the parameters of the potential in general position,
the quantum curve is smooth (i.e., non-degenerate), and
has the maximal possible number of the disconnected
closed contours (``real ovals") in the real section.
In the latter case, the involution is known to divide the
complex curve into two ``halves", thus making the
Schottky double construction applicable.

Before discussing general properties of the curve, we illustrate
its appearance in the semiclassical limit.

\section{Semiclassical curve}\la{SCurve}

By semiclassical case we mean the limit $N \to \infty$, $\hbar \to 0$, while
   the time $t=\hbar N$,
and the potential remain fixed. In this case, all eigenvalues are distributed
   continuously within
droplets with sharp boundaries.
The  boundary of the droplets is a real section of the {\it
classical curve}.

\subsection{Saddle point equation and Riemann-Hilbert
problem}\la{Saddle}In the semiclassical limit,
the distribution
(\ref{mean})  peaks at the minimum of the function
$W(z)-\varphi(z)$,
where
$$
\varphi(z)= -\hbar\sum_{i=1}^N \log|z-z_i|^2
$$
is the Newton potential,
logarithmic in 2-D, of the domain (the support of eigenvalues).
In the semiclassical limit, the potential $\varphi$ is
continuous. One
writes the extremum condition in the form
\begin{equation}\label{s}
\p_z\left(\varphi - W\right)=
\p_{\bar z}\left(\varphi - W\right) =0\,,
\quad \quad \mbox{at}\quad z=z_i.
\end{equation}
The points where this
  equations holds are the most probable positions of
eigenvalues $z_i$.

In addition, if one assumes that $z_i$ have a continuous distribution
(a strong assumption), i.e.,
that
the density
\begin{equation}\label{rho}
\rho_N(z)=\hbar\sum_{i=1}^N\delta(z-z_i)=- \,
\frac{1}{4 \pi} \Delta\varphi
\end{equation}
is a smooth positive function in some domain $D$, and zero outside
this domain, then
it follows from (\ref{s}) that the density is $-\frac{1}{4 \pi}
\Delta W$ inside the domain, if $-\Delta W$ is not singular and
positive inside the support of eigenvalues.

The shape of the domain  is found in two steps.
First, one finds the potential $\varphi$. Then, having the potential,
one has to restore  the shape of the
domain. The second step
is the {\it inverse potential problem}.

Under this assumption,  $\varphi$ is  a harmonic function
in the exterior of the support of eigenvalues. Then
equation (\ref{s}) reads as a Riemann-Hilbert problem:
\begin{description}
\item[(i)]The boundary value of the analytic  function $\p_z\varphi$ in
     the exterior of the domain $D$
is $\p_z W$;
\item[(ii)]
The function $\p\varphi$ behaves at infinity as $-\hbar Nz^{-1}=-t/z$
(we assume that there are no eigenvalues around infinity).
\end{description}
These conditions uniquely specify the domain if it is simply connected.
   In this case, $\pi\hbar N$ is the area of the domain.
Multiply-connected domains are also possible solutions. The number of
disconnected droplets
   cannot exceed the number
of poles of  $A(z)$ plus one. To find them one needs
additional data.
For example, one may
fix the filling factors -- the number of eigenvalues
that each droplet contains.
Obviously, all the filling factors are positive and their sum is $N$.
Domains constructed this way, provided
$A(z)$ is a rational function, belong to a special class of
   {\it algebraic domains} \cite{Ahar-Shap}.

An important property of the Riemann-Hilbert problem is that the function
      $\p\varphi(z)$, with $z$ inside the domain, does not
depend on  $N$, but rather on the potential. The former does not
change during the growth.

At this point,  we specify the potential to be of the form
(\ref{potential}). Then the boundary
value of the analytic function $\p\varphi(z)$ is
       $\p W=-\bar z + A(z)$. The density of eigenvalues
is uniform, and the filling factors  fix
the areas of the droplets
to be  $\pi\hbar\nu_\alpha$.

The potential $\varphi(z)$ can be constructed with the help of
the Schwarz function.

\subsection{The Schwarz function and the inverse potential problem}
       Given a closed (in general a multiply-connected) domain
$D$ in the plane,
the {\it Schwarz function} $S(z)$
is an analytic function
in some neighborhood of the boundary of the domain with the
value  $\bar z$
on the boundary \cite{Alhfors50}:
\begin{equation}\label{Schwarz}
S(z)=\bar z \;\;\;\;\;
\mbox{on the contour}.
\end{equation}
The involution property of the Schwarz
function follows from the definition (cf. (\ref{anti11})):
\begin{equation} \nonumber
\bar S(S(z))=z.
\end{equation}
Let us represent the Schwarz function in the form
$S(z)=S_{+}(z)+S_{-}(z)$, where $S_{+}$ (respectively,
$S_{-}$) is analytic inside (respectively, outside) $D$,
with the condition that $S_{-}(\infty )=0$.
Choose $S_+(z)$ to be equal to $A(z)$,
\be\la{21}
A(z)=\frac{1}{2\pi i}
\oint_{\partial D}\frac{\bar\zeta d\zeta}{z-\zeta}, \quad z\in D.
\ee
Then the function $S_-(z)=S(z)-A(z)$, being analytic outside $D$,
     has the boundary value
     $\bar z-A(z)$. We conclude that
$A(z)-S(z)=\p\varphi(z)$.

\begin{figure} \begin{center}
      \includegraphics*[width=5cm]{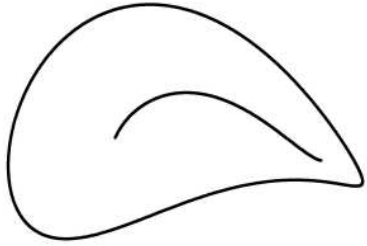}
      \caption{\label{dropletandacut} The boundary
of a droplet and a cut of the
Schwarz function
located inside the droplet. }
      \end{center} \end{figure}

Therefore, the problem of the previous section reads:
given a potential $A(z)$, find the Schwarz function
\be\la{123}
S(z)=A(z)+S_-(z)
\ee
   whose poles and residues outside of the
support of eigenvalues are given by
    the  poles and residues of $A(z)$ and
such that $S_{-}(z) \, \to \, \hbar N/z$ as
     $z\to\infty$.
If the Schwarz function
is known, Eq. (\ref{Schwarz}) determines the domain. Inside the droplet
$S_-(z)$, and therefore, $S(z)$ have cuts
(Fig.~\ref{dropletandacut}).

A note is in order.
In Sec.~\ref{Saddle} we assumed that the density of
eigenvalues is a continuous function
everywhere except for the boundary of $D$. As a consequence,
$\varphi$ is a harmonic function, $\p\varphi$  and $S_-(z)$ are  analytic
functions
outside $D$, and the Schwarz function
outside the droplets has the same poles as $A(z)$.
As a result, the complement to the
support of eigenvalues appears to be an
algebraic domain.

It seems that this assumption is  too restrictive.  Moreover,
   the definition of the curve through the linear differential equations
   for the  wave function of Sec. \ref{Curve1} does not forbid  the
physical branch of the Schwarz function
to have branching points and cuts outside physical droplets, as shown
   in Figs.~\ref{Joukowsky}, \ref{hypotrochoid}.
The semiclassical
wave function still peaks on the real section of the curve (\ref{S1}),
regardless of
whether there are cuts outside the droplets, or if those cuts shrink
  to points.
In other words, a solution with cuts outside
physical droplets  still has a
local maximum on the boundary of the physical droplets and it is stable.
It becomes unstable, however,
   in a vicinity of cuts along a normal direction to a cut.

These arguments suggest
that one can safely allow
$\p\varphi (z)$  to have branching points, i.e., the Schwarz function is
allowed to have
cuts outside physical droplets. Then we seek for solutions of the
Riemann-Hilbert problem
of  Sec. \ref{Saddle} in a wider class of functions.
This leads to a wider class of domains (not algebraic). However, in this case,
one is no longer able to
treat eigenvalues as continuously distributed.
We will come back to this issue in
the next section.

\section {The Schwarz function and its Riemann surface. The Schottky
double.}\label{S}

The Schwarz function describes more than just the boundary of
clusters of eigenvalues.
Together with other sheets
it defines a Riemann surface.
If the   potential $A(z)$ is meromorphic,
the Schwarz function is an algebraic function. It
satisfies a polynomial equation $f(z, S(z))=0$.

\begin{figure} \begin{center}
      \includegraphics*[width=5cm]{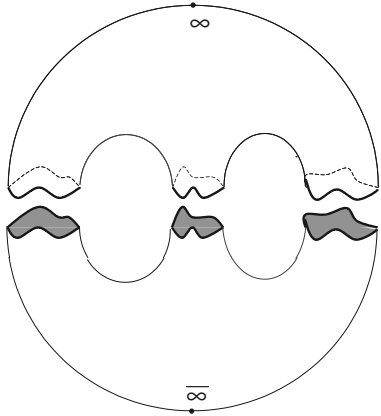}
      \caption{\label{Schottky} The Schottky double. A Riemann surface with
boundaries
     along the droplets (a front side) is
  glued to its mirror image (a back side).}
      \end{center} \end{figure}

The function  $f(z,\tilde z)$, where $z$
and $\tilde z$ are treated as two independent complex arguments, defines
a Riemann surface with antiholomorphic involution (\ref{anti1}).
If the involution divides the surface into two disconnected
parts, as explained above,
the Riemann surface is
       the {\it Schottky double} \cite{C,SS} of
one of these parts.

There are two complementary ways to describe
this surface. One is through the algebraic covering
(\ref{h1}, \ref{ah1}). Among $d$ sheets we distinguish
a {\it physical} sheet.
The physical sheet is
selected by the condition that  the differential $S(z)dz$ has the same
poles and residues  as
the differential of the potential $A(z)dz$ (as in (\ref{123})).
It may happen that the condition
$\bar z=S^{(i)}(z)$ defines a planar curve (or
several curves, or a set of isolated points)
for branches
other than the physical one. We refer to the interior of these
   planar curves as {\it virtual} (or
unphysical) droplets
     situated on sheets other than physical.

Another way emphasizes the antiholomorphic involution. Consider a
meromorphic function $h(z)$ defined
on a Riemann surface with boundaries. We call this surface the front side.
The Schwarz reflection principle extends   any meromorphic function on
the front side
     to a meromorphic function on the Riemann surface without
boundaries. This is done
     by adding another copy of the Riemann surface with
boundaries (a back side),
glued to the front side along the boundaries, Fig~\ref{Schottky}.
The value of the
     function $h$ on the mirror point on the back side
is $h(\overline{S(z)})$.
     The copies are glued along the boundaries:
$h(z)=h(\overline{S(z)})$
if the point $z$ belongs to the boundary.
The same extension rule applies to differentials.
Having a meromorphic differential $h(z)dz$ on the
front side, one extends it to  a
meromorphic differential
$h(\overline{S( z)})d\overline{S(z)}$ on the back side.

This definition can be applied to  the Schwarz function itself.
We say that the Schwarz
     function on the double is $S(z)$ if the point is
on the front side, and
     $\bar z$ if the point belongs to the back side
(here we understand $S(z)$
as a function defined on the complex curve, not just on the physical sheet).

\begin{figure} \begin{center}
      \includegraphics*[width=5cm]{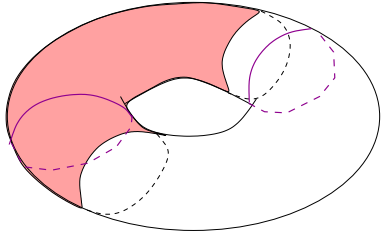}
      \caption{\label{torus}Physical and unphysical droplets on a  torus.
The physical sheet (shaded) meets the unphysical sheet
along the cuts. The cut situated inside the
unphysical droplet appears on the physical sheet.
The boundaries of the droplets (physical and virtual)
belong to different sheets.
This torus is the Riemann surface corresponding to the ensemble
with the potential $V(z) = - \alpha \log (1- z/ \beta) - \gamma z $.}
\end{center} \end{figure}

The number of sheets of the curve is the number of poles (counted
with their multiplicity)
of the function $A(z)$ plus one.
Indeed, according to (\ref{123}) poles of $A$   are  poles of
     the Schwarz function on the front side of the double. On the
back side, there is also a pole at
infinity. Since $S(z=\infty)=A(\infty)$, we have
$\bar S(\bar z=A(\infty))=\infty$.
Therefore, the factor $a(z)$ is a polynomial
with zeros at the poles of $A(z)$
and at $\overline{A(\infty)}$, and
$$d\equiv\mbox{number of sheets} =
\mbox{number of poles of $A$ + 1}.$$
The front and  back sides meet at planar curves $\bar z=S(z)$.
These curves are
  boundaries of the droplets. We repeat  that not all droplets
are physical.
Some of them may belong to unphysical sheets, Fig.~\ref{torus}.

Boundaries of droplets, physical and virtual, form a subset of the
$\bf a$-cycles on the curve. Their number
 cannot exceed the genus of the
curve plus one:
$$\mbox{number of droplets} \le g+1.$$

The sheets meet along cuts located inside droplets.
The cuts that belong to physical droplets show up on  unphysical sheets.
On the other hand, some
     cuts show up on the physical sheet (Fig~\ref{torus}).
They correspond to droplets situated on
     unphysical sheets.

The Riemann-Hurwitz theorem computes the genus of the curve as
$$ g=\mbox{half
the number of
branching points} - d+1. $$

With  the help of the Stokes formula, the numbers $\{ \nu_{\alpha} \}$
are identified with  areas of
the droplets:
$|\nu_a|=\frac{1}{2\pi  \hbar}\int_{D_a} d^2z$.
   For a nondegenerate curve,
   these numbers are not necessarily positive. Negative
numbers correspond to droplets located on unphysical sheets. In this
case, $\{\nu_a\}$
do not correspond to the number of eigenvalues located inside each droplet,
as it is the case for algebraic domains, when all
filling numbers are positive.

A comment is in order. In the Hermitian or unitary
one-matrix ensembles, the
construction of the Riemann
surface is different (Refrs. \cite{DV03,Chekhov,David,Kostov}). In
the Hermitian random matrix
ensemble, for example,
   eigenvalues are real. The support of eigenvalues  is a set of
disjoint intervals on the real line.
       In this case,  branch cuts are identical to the support of
eigenvalues. The Riemann surface
is constructed by gluing $g$ sheets along  $g+1$ cuts \cite{Kostov}.
Conversely, in the Schottky double the cuts generally do not touch
boundaries of the droplets (Fig.~\ref{dropletandacut}).
Complex curves for the Hermitian or unitary ensembles are simpler.
They are always hyperelliptic.

\subsection{Degeneration of the spectral curve}
Degeneration of the complex curve gives the most interesting physical
aspects of growth.
There are several levels of degeneration. We briefly discuss them below.

\subsubsection{Algebraic domains and double points}\la{alg}
A special  case occurs when the Schwarz function on the physical sheet
is meromorphic. It
     has no other singularities than poles of $A$.
This is the case of algebraic domains \cite{Ahar-Shap}.
They appear in the
semiclassical case (see Sec.~\ref{SCurve}).
This situation occurs if  cuts
     on the physical sheet, situated outside physical droplets, shrink
     to  points, i.e., two or more
branching points merge. Then  the physical sheet meets other
     sheets along cuts situated inside physical droplets only and also at
some points on their exterior ({\it double points}) (see figure captions for
  Figs.~\ref{Joukowsky}, \ref{hypotrochoid}).
     In this case the Riemann surface degenerates. The genus
is given by the number of physical
droplets only. The filling factors (\ref{18}) are all positive.

\begin{figure} \begin{center}
      \includegraphics*[width=5cm]{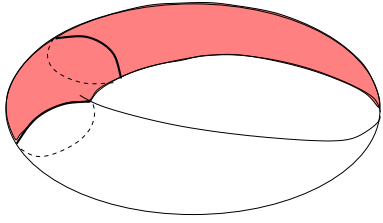}
      \caption{\label{degtorus}Degenerate torus corresponds to the
algebraic domain for the Joukowsky map.}
\end{center} \end{figure}

In the case of algebraic domains, the physical branch of the
Schwarz function
     is a well-defined
meromorphic function. Analytic continuations of $\bar z$
     from different disconnected parts of the boundary
give the same result. In this case, the
     Schwarz function can be written through the Cauchy transform of the
physical droplets:
\begin{equation}\label{28}
S(z)=A(z)+\frac{1}{\pi}\int_{D}\frac{d^2 \zeta}{z-\zeta}.
\end{equation}

\begin{figure} \begin{center}
      \includegraphics*[width=5cm]{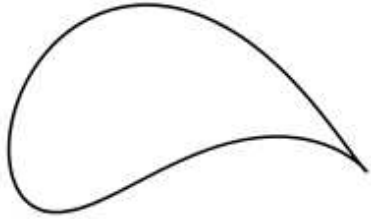}
      \caption{\label{criticalwing}Critically degenerate Joukowsky map.}
\end{center} \end{figure}

Although algebraic domains occur in physical problems
such as Laplacian growth,
their semiclassical evolution is limited.
Almost all algebraic domains will be broken in a  growth process.
Within a finite
     time (the area of the domain) they degenerate further into
critical curves
(Fig.~\ref{evolutionhyp}, \ref{criticalwing}).
The Gaussian potential (the Ginibre-Girko ensemble),
which leads to a single
droplet of the form of an ellipse is a known exception.

\subsubsection{Critical degenerate curves}

Algebraic domains appear as
a result of merging of simple branching points on the physical
sheet. The double points are located
outside physical droplets.
Remaining branching points belong to
the interior of
physical droplets. Initially, they survive
in the degeneration process.
However, as known in the theory of Laplacian growth,
   the process necessarily leads to a further
degeneration. Sooner  or later, at least one of the
interior branching points  merges with one of the double points in
the exterior.  Curves degenerated in this
manner are called
{\it critical}. For the genus one and three
this degeneration is discussed below.
The critical degenerations
are depicted, respectively, in Fig.~\ref{criticalwing} and Fig.~\ref{evolutionhyp}.

Since interior branching points
can only merge with exterior branching points
on the boundary of the
droplet, the boundary develops a cusp, characterized by a pair $p, q$
of mutually prime integers. In local coordinates
around such a cusp, the curve looks like $x^p\sim y^q$.

The fact that the growth of algebraic domains always leads to critical
curves is known
in the theory of Laplacian growth (see e.g., \cite{review})  as
finite time singularities. It is also
known in the Hermitian one- and two-matrix
models as
critical points (intensively
studied in 2-D-gravity \cite{gravity}).

The degeneration process seems to be a feature of the semiclassical
approximation. Curves treated beyond this approximation never
degenerate.

\subsubsection{Simply-connected domains. Conformal maps.}\la{C}
The case when the complement to the support of eigenvalues is a
simply-connected algebraic domain
in the extended (i.e. including $\infty$) complex plane
is particularly important.
All the filling factors are zero except
one, which is equal to $N$, so there is only one droplet $D$.
In this case, the Schottky double of the
exterior of the droplet is a Riemann sphere.

The exterior domain can be
conformally and univalently mapped onto the exterior of the unit disc.
We call this map $w(z)$, and its inverse
$z(w)$.
For algebraic domains, the inverse map is
a rational function of $w$.
Choosing the normalization such that
$z(\infty )=\infty$, we represent the
inverse map
by a half-infinite Laurent polynomial
\begin{equation}\la{291}
z(w)=r w+\sum_{k>0}u_{k} w^{-k},\quad |w|\geq 1.
\end{equation}
The coefficient $r$, called
the external conformal radius of the
domain $D$, is chosen to be real positive.
The  Schwarz reflection of the inverse map is
\begin{equation}\la{33}
\bar z(w^{-1})=r w^{-1}+\sum_{k>0}{\bar u^{(k)}}{w^k},\quad |w|\geq 1.
\end{equation}
Given a point $z$, there are $d$ values of $w$ solving the
equation (\ref{291}), where $d$ is the number of poles
of the rational function $z(w)$. Among them,
only one solution corresponds to
the conformal map. It is the solution
such that $w\to z/r$ as $z\to\infty$.
The global coordinate $w$ on the Riemann sphere
provides a uniformization of
the Schottky double. In this coordinate, the involution
reads $w\to 1/\bar w$.
The Schwarz function is
\be\la{ss}S(z)=\bar z(w^{-1}(z)).
\ee

Choose a regular  point  $\xi_0$ of the potential inside the
droplet and expand
\begin{equation}\la{}
V(z)=\sum_{k>0} t_k (z-\xi_0)^{k}.
\end{equation}
The coefficients $t_k$ have a simple geometric
interpretation. It follows from (\ref{21}) that  they are
{\it harmonic moments}
of the exterior of the droplet
(with respect to the point $\xi_0$):
\begin{equation}\label{im2}
t_k =-\frac{1}{\pi k}\int_{{\bf C}\setminus D}(z-\xi_0)^{-k}d^2 z.
\end{equation}
We also mention the area formula
\be\la{area}
t=r^2-\sum_{k\geq 1}k|u^{(k)}|^2.
\ee

If  the vector potential $A(z)$ is
a polynomial of degree $d-1$,  then
the inverse map $z(w)$ is a (Laurent) polynomial:
$u^{(k)}=0,\quad k>d-1$. If $A(z)$
   is a rational function with $d-1$ simple poles
in the finite part of the complex plane,
the map $z(w)$ is a rational function
with $d-1$ simple poles and one extra simple pole at infinity.

\subsection{The generating differential}\la{Diff}

The meromorphic differential
\be\la{Omega}
d\Omega=S(z)dz
\ee
plays an important role.
It is called generating
differential \cite{mkwz,kkmwz}.
According to (\ref{123}), on the physical sheet
it has the same poles and residues
as the differential $Adz$:
\be\la{1234}
d\Omega=Adz+S_-(z)dz.
\ee

Below we use the following properties of the
generating differential.
\begin{itemize}
\item [(i)]
The periods over $\bf a$-cycles
(boundaries of the droplets) are
purely imaginary, and are integer multiples
of $2\pi i$. They compute   areas
of the droplets (the filling factors (\ref{18})):
$$
\nu_{\alpha}=\frac{1}{2\pi i \hbar}\oint_{{\bf a}_{\alpha}}
d\Omega.
$$
The filling factors of physical droplets
(belonging to the physical sheet)
are positive.
\item[(ii)]
The real part of the integral of the differential
$(\bar z-S(z))dz$ from some fixed
point $\xi_0$ to a point on the boundary
of  a droplet (a point on a $\bf a$-cycle)
     has the same value for all points of the boundary:
\be \nonumber
\phi_\alpha=-|z|^2+2 {\mathcal R}e\int_{\xi_0}^z d\Omega=
\mbox{const,\quad\quad for all}\quad z\in {\bf a}_\alpha.
\ee
This quantity does not depend on $z$,
but does depend on $\xi_0$ unless $\xi_0$ is on the boundary.
However, the difference $\phi_{\alpha}-\phi_{\beta}$
depends on the ${\bf a}$-cycles only.
It is equal to a $\bf b$-period of the differential
$d\Omega$:
     \be\la{32}
\phi_\alpha-\phi_\beta=\oint_{{\bf b}_{\alpha\beta}}d\Omega,
\ee
where ${\bf b}_{\alpha\beta}$ is a
cycle connecting ${\bf a}_{\alpha}$ and
${\bf a}_{\beta}$ cycles.
\end{itemize}
For proofs and more details, see Ref. \cite{mkwz}.

Periods over $\bf b$-cycles $\phi_\alpha-\phi_0$  play a role of
chemical potentials for the filling factors.
Here $0$ denotes a chosen reference droplet.
One can use chemical potentials to
characterize evolution of the curve instead
of filling factors.

\section{Examples}\label{Example}

\subsection{Genus zero -- ellipse}\la{classicalellipse}

The potential is Gaussian, $V(z)= t_2 z^2$,
$2|t_2| <1$, $A(z)=2t_2 z$. It  has a simple
pole at infinity.
The number of sheets
of the Schwarz function is two.
At infinity,  $S'(z)$ is finite. Therefore,
$S'(z)$ takes every value
      twice. It must have two zeros at most, i.e.,
there are two branching points.
The Riemann-Hurwitz theorem says that genus is zero.
The curve with two branching points and two sheets
has the form
$f(z,\bar z)=z\bar z+k_1z^2+\bar k_1\bar z^2+k_2=0$. Conditions that
$S(z)\sim 2t_2z+n\hbar z^{-1}+\dots$ at $z\to\infty$  for the
physical branch and the unitarity
condition  (\ref{anti11}) determine the curve in full:
\be\la{curveellipse}
f_n(z,\bar z)=z\bar z - ( t_2z^2+ \bar t_2\bar z^2)
\frac{2}{1+4|t_2|^2}-n\hbar\frac{1-4|t_2|^2}{1+4|t_2|^2}=0.
\ee
The physical branch
$$
S^{(1)}(z)=2t_2 z+\frac{(2\bar
t_2)^{-1}-2t_2}{2}z
\left (1-\sqrt{1-\frac{2}{(2\bar
t_2)^{-1}-2t_2}\frac{2n\hbar}{z^2}}\right )
$$
is the Schwarz function of ellipse.
     The second branch  $S^{(2)}(z)\to (2\bar t_2)^{-1}z$
does not correspond to any real curve.
The droplet (see Fig.~\ref{ellipse}) is an ellipse with the
quadrupole moment $2|t_2|$ and area $\pi t=\pi n\hbar$.

Recurrence relations
Eqs. (\ref{M11}, \ref{M}) truncate
\be\la{59}
z\psi_n=r_n\psi_{n+1}+u_n\psi_{n-1},\quad\quad
(L^{\dag}\psi)_n=r_{n-1}\psi_{n-1}+\bar u_{n+1}\psi_{n+1}.
\ee
Equations (\ref{M1}, \ref{Area}) imply
\be\la{58}
2t_2=\frac{\bar u_{n+1}}{r_n},\quad n\hbar=r_n^2-|u_{n+1}|^2,
\ee
where the second equation is an analog of the area formula (\ref{area}).
Together they give a growth
equation for the quantum analog of conformal radius
$n\hbar=(1-|2t_2|^2)r_{n}^2$.

The operator $L^{\dag}$ can be cast in the form of a $2\times 2$
matrix (the number of sheets of the curve). Writing (\ref{59})  for
$n+1$, we express $\chi_{n-1}$ and
$\chi_{n+2}$ through $\underline\chi_n=(\chi_n,\,\chi_{n+1})$ from the
first pair,  and substituting them to the second pair we obtain
\be\la{601}
{\mathcal L}_n \underline\chi_n=
\left (
\begin{array}{cc}
     z\frac{ r_{n-1}}{u_n} & \bar u_{n+1} - \frac{r_n r_{n-1}}{u_n}  \\
     r_n- \frac{\bar u_{n+2}u_{n+1}}{r_{n+1}} &
\frac{\bar u_{n+2}}{r_{n+1}}z
\end{array}
\right)\underline\chi_n.
\ee
Formally, one can say that  the recurrence  relation (\ref{59})
represents the shift operator as the $2\times 2$ matrix
$$
{\mathcal W}_{n-1}=
\left (
\begin{array}{cc}
0 & 1  \\
-\frac{u_n}{r_n}  &
\frac{z}{r_n}
\end{array}
\right).
$$
The relation ${\mathcal L}_n =\left (
\begin{array}{cc}
r_{n-1} & 0 \\
0  & r_{n}
\end{array}
\right){\mathcal W}^{-1}_{n-1}+\left (\begin{array}{cc}
\bar u_{n+1} & 0 \\
0  & \bar u_{n+2}
\end{array}
\right){\mathcal W}_n
$ then
yields (\ref{601}).
With the help of (\ref{58}),  ${\mathcal L}_n$ reads
$$
{\mathcal L}_n (z)=
\left (
\begin{array}{cc}
z(2\bar t_2)^{-1} & -r_n(2\bar t_2)^{-1}(1-4|t_2|^2)  \\
r_n(1-4|t_2|^2)  & 2t_2 z
\end{array}
\right).
$$
Computing the determinant $\det ({\mathcal L}_n (z) - \bar z)$, we obtain the
curve (\ref{curveellipse}), already
determined  by the singularities.

The recurrence relation can be used to generate the
Hermite polynomials.
The model with potential $V(z) = t_2 z^2 + Q\log z$ can
also be solved explicitly (Ref. \cite{Akemann02}). It
generates  the Laguerre polynomials. These are  the cases where
biorthogonal polynomials are classical orthogonal
polynomials.

Classical polynomials  correspond to a
genus  zero curve. Higher genus curves generate more
complicated, but more interesting polynomials.

The classical limit of the recurrence relations
gives a conformal map of the exterior
of the unit disk to the exterior of the ellipse,
$z(w)=rw+\frac{u}{w}$.
The  double-valued function
$
w_{1,2}(z) = \frac{1}{2r} \left [
z \pm \sqrt{(z -z_1)(z-z_2) } \right ],\quad z_{1,2}=\pm 2\sqrt{ur}
$
becomes single-valued on a two-sheet covering
of the $z$-sphere plane. The branch $w_1(z)$ is
such that $w_1\to\infty$ as $z\to\infty$.
It defines the inverse map from the ellipse exterior
onto the exterior of the unit disk in the $w$-plane.
The function
$
\bar z(w^{-1})=rw^{-1} + \bar u w
$
      is a meromorphic function of
$w$ with one simple pole at $w=0$. Treated as  functions of $z$, the
two sheets $S^{(1,2)}(z)=\bar z(w_{1,2}^{-1}(z))$
have two critical points $z_{1,2}$ and constitute a sphere.
Parameters of the map are related to the parameters of the potential
through the classical limit
of (\ref{58}): $t_2=\frac{\bar
u}{2r},\quad t=r^2-|u|^2$.
The general formula for the semiclassical behavior of the wave function
(\ref{sp1})
   gives the known asymptotics of the
Hermite polynomials \cite{BE}.

The cut inside the droplet may shrink to a point.
In this case only one (physical) sheet remains,
and the Schwarz function becomes a rational
function with one simple pole.
The wave functions are monomials.
The droplet is a disk.

A more interesting degeneration
occurs at the critical value of $|t_2|=\frac{1}{2}$,
when the two branching points
reach the boundary of the droplet.
The ellipse degenerates into the cut $[z_1,\,z_2]$.

Beyond the critical value
($|t_2|> \frac{1}{2}$), the ellipse appears again, but on
the unphysical sheet. In this case, there is no physical droplet,
but there is a virtual droplet. At this value of $t_2$, the matrix
integral diverges,
but a solution of the differential equation (\ref{M2}) exists. It
is not a polynomial,
however.

Below we use a more direct method to find the quantum curve
bypassing the matrix
representation of $L^{\dag}$. The method reflects the triangular
structure of $L^{\dag}$,
and closely resembles the procedure used in the classical case.
We will use this method for the following, more complicated, examples.

Apply (\ref{59}) to an eigenvector $c_n$ of the matrix
    ${\mathcal L}_n (z)$ with an eigenvalue $\tilde z$. The equations
\be \nonumber
\left
\{
\begin{array}{lcl}
z c_{n}& = & r
_n c_{n+1} + u_n c_{n-1} \\
\tilde z c_{n} & = & \bar u_{n+1} c_{n+1} + r_{n-1} c_{n-1}
\end{array}
\right.
\ee
are compatible if the point $(z, \tilde z)$
belongs to the curve. To obtain a
compatibility  condition,
we express $c_{n+1}$ and $c_{n-1}$ in terms of $c_n$. The
results must differ
by a shift $n\to n+2$
\be \nonumber
    \det
\Big |
\begin{array}{cc}
z & u_n \\
\bar z & r_{n-1}
\end{array}
\Big |\det
\Big |
\begin{array}{cc}
r_{n+1} & z \\
\bar u_{n+2} & \bar z
\end{array}
\Big |= d_nd_{n+1},
\quad
d_n = \det
\Big |
\begin{array}{cc}
r_n & u_n \\
\bar u_{n+1} & r_{n-1}
\end{array}
\Big |.
\ee
This gives the curve (\ref{curveellipse}).

\subsection{Genus one -- an aircraft wing}\la{J}

The potential is
$V(z) = - \alpha \log (1- z/\beta) -
\gamma z,\quad A(z)=-\frac{\alpha}{z-\beta}-\gamma$.
There is one pole at $z=\beta$  on the first (physical) sheet.
At $z=\infty$ on the first sheet
$S(z)\to-\gamma+\frac{n\hbar-\alpha}{z}$.
Therefore, the Schwarz function has another pole
at the point $-\bar\gamma$
on another sheet.  All the poles are simple.
According to the general arguments of
    Sec.~\ref{S},  the number of sheets is 2,
the number of branching points is 4. The genus is 1.
The curve has the form
$$
f(z, \bar z) = z^2 \bar z^2 +k_1
z^2\bar z +\bar k_1 z\bar z^2  + k_2 z^2 + \bar k_2 \bar z^2 +
k_3 z\bar z
+k_4z+
\bar k_4\bar z +h =0.
$$
The points at infinity  and  $-\bar\gamma$ belong to
the second
sheet of the algebraic covering. Summing up,
\[ S(z)=
\left \{\begin{array}{rll}
-\frac{\alpha}{z-\beta}& \mbox{as} & z\to\beta_1, \\
(-\gamma+\frac{n\hbar-\alpha}{z} ) & \mbox{as} & z\to\infty_1, \\
\frac{n\hbar-\bar\alpha}{z+\bar\gamma}& \mbox{as}
& z\to -\bar\gamma_2, \\
(\bar\beta-\frac{\bar\alpha}{z})& \mbox{as} & z\to\infty_2.
\end{array}
\right. \]
where, by 1 and 2  we indicate the sheets.

Poles and residues of the Schwarz function
determine   all the coefficients of the curve
$f(z,\bar z)=a(z) (\bar z-S^{(1)}(z))(\bar z-S^{(2)}(z))=
\overline{a(z)} (z-\bar S^{(1)}(\bar z))(z-\bar
S^{(2)}(\bar z))$ except one.
The behavior at $\infty$ of $z, \bar z$
gives $k_1=\gamma -\bar \beta, \, k_2=-\gamma \bar \beta$.
Hereafter we choose the origin by
setting  $\gamma=0$.
The equation of the curve then reads $f_n(z, \bar z) = 0$, where
$f_n(z, \bar z) $ is given by
\be\la{8}
z^2 \bar z^2 -
z^2\bar z \bar\beta  -z\bar z^2\beta  +
\left ( |\bar\beta|^2 +\alpha+\bar\alpha-n\hbar
\right ) z\bar z
+z\bar\beta(n\hbar-\alpha)+
\bar z\beta(n\hbar-\bar\alpha)
+h_n
\ee
The free term $h_n$ is to be determined by
filling factors of the two
droplets $\nu_1$ and $\nu_2=n-\nu_1$.
A detailed analysis shows
that the droplets belong to different
     sheets (Fig.~\ref{torus}).
Therefore, $\nu_2$ is negative.

A boundary of a physical  droplet is given by the
equation
$\bar z=S^{(1)}(z)$ (Fig.~\ref{Joukowsky}).
The second droplet
belongs to the unphysical sheet. Its
boundary is given by
     $\bar z=S^{(2)}(z)$. The explicit form of both branches is
$$S^{(1,2)}=\frac{1}{2}\bar\beta-
\frac{\beta(n\hbar-\bar\alpha)+(\alpha+\bar\alpha-n\hbar)z\mp
\sqrt{(z-z_1)(z-z_2)(z-z_3)(z-z_4)}}{2(z-\beta)z},$$
where the branching points $z_i$ depend on $h_n$.

If the filling factor of the physical droplet is equal to $n$, the cut
inside the unphysical droplet is of the order of
   $\sqrt{\hbar}$. Although it never vanishes, it
  shrinks to a double point
$z_3=z_4=z_*$ in a semiclassical limit.
The sheets meet at the double
point $z_*$ rather than along the cut:
$\sqrt{(z-z_1)(z-z_2)(z-z_3)(z-z_4)}\to
(z-z_*)\sqrt{(z-z_1)(z-z_2)}$.
In this case, genus of the curve reduces to zero and
the exterior of the physical droplet becomes an algebraic domain.
This condition determines $h$,
and also the position of the double point (Fig.~\ref{degtorus}).
The double point is a saddle point for the
level curves of  $f(z,\bar z)$.
If all the parameters are real, the double point is stable in
$x$-direction and unstable in
$y$-direction. A saddle point is a signature of a virtual droplet
\cite{DV03}.

If this solution is chosen,
the exterior of the physical droplet can be mapped
to the exterior of the unit disk by the Joukowsky map
\be\la{222}
z(w)=rw+u_0 + \frac{u}{w-a},\quad |w|>1,\quad |a|<1.
\ee
The inverse map
is given by the branch $w_1(z)$ (such that $w_1\to\infty$ as
$z\to\infty$)
of the double valued function
$$
w_{1,2}(z) = \frac{1}{2r} \left [
z-u_0 + ar \pm \sqrt{(z -z_1)(z-z_2) } \right ],\quad
z_{1,2}=u_0+ar\mp 2\sqrt{r(u+au_0)}.
$$

     The function
\be\la{23}
\bar z(w^{-1})=rw^{-1}+\bar u_0 + \frac{\bar u}{w^{-1}-\bar a}
\ee
     is a meromorphic function of
$w$ with two simple poles at $w=0$ and $w=\bar a^{-1}$. Treated as a
function of $z$, it  covers
the $z$-plane twice. Two branches of the Schwarz function are
$S^{(1,2)}(z)=\bar z(w_{1,2}^{-1}(z))$. On the physical sheet,
$S^{(1)}(z)=\bar z(w_1(z))$ is the analytic continuation
of $\bar z$ away from the boundary. This function is
meromorphic outside the droplet. Apart from
a cut between the branching points
$z_{1,2}$, the sheets also meet at
the double point $z_*=-\bar\gamma+a^{-1}re^{2i\phi}$,
where $S^{(1)}(z_*)=S^{(2)}(z_*), \phi = \arg (ar+\frac{u\bar a}{1-|a|^2})$.

Analyzing singularities of the Schwarz function,
   one  connects parameters  of
the conformal map with the deformation parameters:
\be\la{27}
\left \{
\begin{array}{rcl}

     \gamma  & = &  \frac{\bar u}{\bar a} - \bar u_0,  \\
n\hbar -\bar\alpha  & = &  r^2 -\frac{ur}{a^2},  \\

     \beta & = &  \frac{r}{\bar a} + u_0  + \frac{u \bar a}{1-|a|^2}
\end{array}\right.
\ee
$$
\mbox{Area of the droplet}\sim n\hbar=r^2-\frac{|u|^2}{(1-|a|^2)^2}.
$$

A critical degeneration occurs when the double point merges with a
branching point located inside the droplet
($z_*=z_2$) to form a triple point $z_{**}$.
This may happen on the boundary only.
At this point, the boundary has a $(2,\,3)$
cusp (Fig.~\ref{criticalwing}).
   In local coordinates, it is
$x^2\sim y^3$. This is a critical point of the conformal map:
$w'(z_{**})=\infty$.
A critical point
inevitably results from the evolution
at some finite critical area.

     A direct way to  obtain the
complex curve from the conformal map is the following. First,
rewrite (\ref{222}) and (\ref{23}) as
\be\la{22}\left
\{
\begin{array}{lcl}
z-u_0+ar & = & rw+a(z+\bar\gamma)w^{-1}\\
\bar z-\bar u_0+\bar a r & = & rw^{-1}+\bar a(\bar z+\gamma)w,
\end{array}
\right.
\ee
and treat $w$ and $1/w$ as independent variables.
Then impose the condition $w\cdot w^{-1} = 1$. One obtains
$$
\left |\det
\left [
\begin{array}{cc}
z- u_0 + ar & a(z+\gamma) \\
\bar z -\bar u_0 + \bar a r & r
\end{array}
\right ]\right |^2
=\left (\det
\left [
\begin{array}{cc}
r & a(z+\bar\gamma) \\
     \bar a( \bar z+\gamma) & r
\end{array}
\right ]\right ) ^2.
$$
This gives the equation of the curve and in particular $h$,
in terms of $u,\, u_0,\,r,\,a$ and eventually through the deformation
parameters $\alpha,\,\beta,\,\gamma$ and $t$.

The semiclassical analysis gives a guidance for the form of the
recurrence relations.
Let us use an ansatz for the $L$-operator, which resembles
   the conformal map (\ref{222}):
$$
L=r_n \hat w +u_{n}^{(0)} + (\hat w -a_n )^{-1} u_n,
$$
so that
\begin{eqnarray}
(\hat w -a_n ) L = (\hat w -a_n ) r_n \hat w +
(\hat w - a_n ) u^{(0)}_n + u_n, \la{701} \\
L^{\dag}(\hat w^{-1} -\bar a_n ) =
\hat w^{-1} r_n (\hat w^{-1} -\bar a_n ) +
\bar u^{(0)}_n (\hat w^{-1} -\bar a_n ) + \bar u_n \, \la{7111},
\end{eqnarray}
where $\hat w$ is the shift operator $n\to n+1$.

Now we follow the procedure of the previous section. Since the
potential has only
    one pole,  ${\mathcal L}_n$ can be cast into
    $2\times 2$ matrix form. Let us
    apply the lines (\ref{701}, \ref{7111})
    to an eigenvector $(c_n, c_{n+1})$ of a yet unknown operator
${\mathcal L}_n$, and set
   the eigenvalue to be $\tilde z$:
\be\la{94}\left
\{
\begin{array}{ccc}
(z+r_{n-1} a_{n-1}-u^{(0)}_{n})c_{n} & = & r_{n} c_{n+1} +
a_{n-1}(z+\bar \gamma_{n-1})c_{n-1} \\
(\tilde z +  r_{n}\bar a_{n} -
\bar u^{(0)}_{n+1} )c_{n} & = &
\bar a_{n+1}(\tilde z + \gamma_{n+1}) c_{n+1}  + r_{n} c_{n-1}.
\end{array}\right.
\ee
We have defined $\bar \gamma_n = \frac{u_n}{a_n} - u^{0}_n$.
The equations are compatible if $c_{n-1}$ and $c_{n+1}$ found
through $c_n$
    differ by the shift $n\to n+2$. We have
\be \nonumber
c_{n+1} = \frac{c_n}{d_n} \det \left | \begin{array}{cc}
z+r_{n-1} a_{n-1} -u^{(0)}_{n}&  a_{n-1}(z+\bar \gamma_{n-1}) \\
\tilde z + r_n\bar a_{n} - \bar u^{(0)}_{n+1} & r_{n}
\end{array} \right | = c_n \frac{\widetilde{{\mathcal D}}_{n}}{d_n},
\ee
\be \nonumber
c_{n-1} = \frac{c_n}{d_n} \det \left | \begin{array}{cc}
r_{n} & z + r_{n-1} a_{n-1} -u^{(0)}_{n}  \\
\bar a_{n+1}( \tilde z +
\gamma_{n+1}) & \tilde z + r_n\bar a_{n} - \bar u^{(0)}_{n+1}
\end{array} \right | = c_n \frac{{\mathcal D}_n}{d_n},
\ee
where
\be \nonumber
d_n =
\det \left | \begin{array}{cc}
r_{n}  & a_{n-1}(z+\bar \gamma_{n-1})   \\
    \bar a_{n+1} (\bar z +\gamma_{n+1}) & r_{n}.
\end{array} \right |
\ee
This  yields  the curve
\be\la{811}
\widetilde{\mathcal D}_n\cdot {\mathcal D}_{n+1}=d_nd_{n+1}.
\ee
Comparing the two forms of the curve
(\ref{8}) and (\ref{811}),
    we obtain the  conservation laws of growth:
\be \nonumber
\gamma =\gamma_n = \frac{\bar u_n}{\bar a_n}-\bar u^{0}_n,
\ee
\be\nonumber
\beta  = \frac{r_n}{\bar a_{n+1}} + u^{(0)}_{n+1} + \frac{u^{(0)}_{n+1}
a_n \bar a_{n+1}}{1-a_{n}\bar a_{n+1}},
\ee
\be \nonumber
     n \hbar -\bar \alpha= r_n r_{n+1} - \frac{r_{n+1}u_{n+1}}{a_{n}a_{n+1}} .
\ee
They are the quantum version of (\ref{27}).

The poles of the vector potential field
determine most elements of the  matrix
${\mathcal L}_n$. We have
\be \nonumber
{\mathcal L}_n
\underline\chi_n=\Big({\bf A}_n+\frac{{\bf B}_n}{z}+
\frac{{\bf C}_n}{z-\beta}\Big)
\underline\chi_n,
\quad\quad \underline\chi_n=(\chi_n,\chi_{n+1})^{t},
\ee
where $\bf {A, B}$ and ${\bf C}$ are $z$-independent $2\times 2$
matrices. Their
elements are
$A_{11}=\bar\beta,\,A_{21}=\bar\beta a_n,\quad A_{12}=A_{22}=0,\quad
B_{11}=t-\bar\alpha,\,B_{12}=
\frac{r_{n-1}r_n}{a_{n-1}},\quad B_{21}=B_{22}=0,
\quad\quad \mbox{tr}{\bf C }=-\alpha,\,\mbox{det}{\bf C}=0,\, \hbar n=
-C_{11}-\bar\beta a_n(B_{12}+C_{12})$.
The undetermined matrix elements are
   found similarly to the case of the ellipse.
We treat the recurrence relation (the first line of (\ref{94}))
   as  a $2\times 2$ representation of the shift operator
$$
{\mathcal W}_n=
\left (
\begin{array}{cc}
0 & 1  \\
-z\frac{a_n}{r_{n+1}}  &
\frac{z+r_na_n-u^{(0)}_{n+1})}{r_{n+1}}
\end{array}
\right)
$$
and compute ${\mathcal L}_n$ from the second line of (\ref{94}). The
matrix elements
of ${\bf C}_n$ appear to be rather cumbersome.

\subsection{Genus three -- hypotrochoid}
The potential is $V(z) = t_3 z^3$. A general
polynomial potential,
including this example,
has been analyzed in Ref. \cite{KM03}. We briefly review
it here to illustrate  features of the curve with multiple poles
of the Schwarz function.
In this case, $A(z)$ has a double pole at infinity. The number of
sheets is three.
     The behavior of the
Schwarz function at infinity is $S(z)\sim 3t_3z^2$ and the
     symmetry of the potential under $2\pi/3$-rotation  restricts the
equation of the curve to the form
\be\la{321}
f_n(z,\bar z)=(z \bar z)^2 - \frac{\bar z^3}{3t_3} - \frac{z^3}{3
\bar t_3} + k_nz \bar z + h_n
=0,
\ee
with two unknown coefficients. Since $S(z)$ has a double pole at infinity
     on the physical sheet,
it has a branching point at infinity on the other sheets.
This is the point where two unphysical sheets meet. Eq.(\ref{321})
suggests that
there are 9 more branching points.
Therefore, genus of the curve  is $g=3$. There are four droplets,
     but only one belongs to the physical sheet. $2\pi/3$ rotations leave
the physical
     droplet invariant and exchange unphysical droplets. Therefore, the
filling factors of
unphysical droplets must be equal. This number, together with the
filling factor of
the physical droplet, determine the coefficients in (\ref{321}).

Let us start from the semiclassical analysis. Assume that the
filling factor of the
physical  droplet is $n$. We then have an algebraic domain
bounded by a hypotrochoid, Fig.~\ref{hypotrochoid}.
Six critical points  sitting outside the
droplet collapse two by two to three double points.
The function
\begin{equation} \label{hypotrochoid1}
z(w) = rw + \frac{u}{w^2}
\end{equation}
maps the exterior of the unit disk to the exterior of the
hypotrochoid. The inverse function
has three branches $w_i(z),\quad i=1,2,3$.
At infinity, the three sheets have the leading terms $w_1(z)\to z/r$,
$w_{2,3}(z) \to \pm (u/z)^{1/2}$. Therefore, the inverse
of the map
(\ref{hypotrochoid1}) is $w_1(z)$.
The branch $S(z) = \bar z(w^{-1}_1(z))$ is the Schwarz function.
      It is a meromorphic function outside the domain with one double
pole at infinity,
$S(z)\sim  3t_3z^2+ (n\hbar)/z$.
This gives $\bar u=3t_3r^2$, while the area
is $n\hbar=r^2-2|u|^2=
r^2-18|t_3|^2 r^4$. The coefficients of the equation of the curve
are determined by
\be\la{341}
k  =  \frac{(1-9|t_3|^2 r^2)(1+18|t_3|^2 r^2)}{9|t_3|^2},
\quad
h  = - \frac{r^2(1-9|t_3|^2r^2)^3}{9|t_3|^2}.
\ee
Four remaining branching points are $z=\infty$ and
three  points solving $d z =
(r-2uw^{-3})d w =0$,  at $w_{*}^3 = 2u/r$. They are situated  inside the
droplet.
Three double points solve
$S^{(i)}(z) = S^{(j)}(z)$.
They are outside the droplet:
$$
z_*^3 = \frac{(r^2-|u|^2)^3}{|u|^2\bar u}.
$$
At a critical degeneration the double points merge on the boundary to
the branching
point inside the droplet, $z_* = z(w_{*}), |w_{*}|=1$
(Fig.~\ref{evolutionhyp}).

Quantum analysis starts from recurrence relations guided by the
classical case.
Eqs. (\ref{M11}, \ref{M}) read
\be\la{591}
z\psi_n=r_n\psi_{n+1}+u_n\psi_{n-2},\quad\quad
(L^{\dag}\psi)_n=r_{n-1}\psi_{n-1}+\bar u_{n+2}\psi_{n+2}.
\ee
Acting on the eigenvector of ${\mathcal L}_n$, we get
\begin{equation*}
\left \{
\begin{array}{ccc}
zc_n & =  & r_n c_{n+1} + u_nc_{n-2} \\
\tilde z c_{n} & = & r_{n-1}c_{n-1} + \bar u_{n+2}c_{n+2}.
\end{array}
\right .
\end{equation*}
Writing the first equation for $n\to n\pm 1$, we express $c_{n+2}$
and $c_{n-1}$
    through $ c_n$ and $c_{n+1}$ to obtain
\be \nonumber
\tilde z c_n  =z \frac{ \bar
  u_{n+2}}{r_{n+1}} c_{n+1} +
\frac{1}{r_{n+1}}( r_{n-1}r_{n+1} -
u_{n+1}\bar u_{n+2}) c_{n-1},
\ee
\be \nonumber
zc_{n}= \left (r_n - \frac{u_n \bar u_{n+1}}{r_{n-2}} \right ) c_{n+1} +
\bar z\frac{ u_n}{r_{n-2}} c_{n-1}.
\ee
Equations are compatible if $\tilde z$ belongs to the curve
\be \nonumber
\det
\left |
\begin{array}{cc}
\left (r_{n+1} - \frac{u_{n+1} \bar u_{n+2}}{r_{n-1}} \right )  & z \\
\frac{z \bar u_{n+3}}{r_{n+2}} & \tilde z
\end{array}
\right |
\cdot
    \det
\left |
\begin{array}{cc}
z & \frac{\tilde z u_n}{r_{n-2}} \\
\tilde z &  \left (r_{n-1} -
\frac{u_{n+1}\bar u_{n+2}}{r_{n+1}} \right )
\end{array}
\right |= d_{n}  d_{n+1},
\ee
where
\be \nonumber
d_n = \det
\left |
\begin{array}{cc}
\left (r_n - \frac{u_n \bar u_{n+1}}{r_{n-2}} \right )
&  \frac{\tilde z u_n}{r_{n-2}} \\
\frac{z \bar u_{n+2}}{r_{n+1}} &
\left (r_{n-1} - \frac{u_{n+1}\bar u_{n+2}}{r_{n+1}}
\right )
\end{array}
\right |.
\ee
Comparing with (\ref{321}), we find the quantum version of
(\ref{341}):
\be \nonumber
k_n = r^2_n \left [ 1-9|t_3|^2 (r^2_{n-1}+r^2_{n+1} ) +
\frac{1}{9|t_3|^2r^2_n} \right ],
\ee
\be \nonumber
h_n = -
\frac{r^2_n}{9|t_3|^2}(1-9|t_3|^2r^2_{n-1})
(1-9|t_3|^2r^2_{n})(1-9|t_3|^2r^2_{n+1}),
\ee
where we used
    the  conservation law and a quantum  analog of the area
formula
\be \nonumber
3t_3 = \frac{\bar u_{n+1}}{r_{n-1}r_{n+1}},
\quad
r_n^2-(|u_{n+2}|^2+
|u_{n+1}|^2)=\hbar n.
\ee
Together, these equations
give a closed equation for $r_{n}^{2}$,
$$
r_n^2(1-|3t_3|^2(r_{n+1}^2+r_{n-1}^2))=\hbar n,
$$
which is a discrete analog
of the Painlev\'e I equation.

The matrix form of $L^{\dag}$ can be obtained in a similar manner.
    It is  a $3\times 3$ matrix acting on the 3-vector
$(\chi_{n},\chi_{n+1},\chi_{n+2})^{{\rm t}}$.
Its analytic structure is
$${\mathcal L}_{n}={\bf A}_n+z{\bf B}_n+z^2 {\bf C}_n,$$
and the matrices ${\bf A},\,{\bf B},\,{\bf C}$ are
$$\displaystyle
{\bf A}_n=\begin{matrix}
\left(
\begin{array}{ccc}
      0 &
0 & \frac{r_n}{r_{n-1}}(3\bar
t_3)^{-1}(1-|3t_3|^2r_{n-1}^2) \\
       r_{n-1}(1-|3t_3|^2r_n^2)& 0 &  0
      \\
0 & r_{n}(1-|3t_3|^2r_{n+1}^2) & 0
\end{array}
\right )
\end{matrix},
$$
$$\displaystyle
{\bf B}_n=\begin{matrix}
\left(
\begin{array}{ccc}
      0 &
\frac{1}{ r_{n-1}}(3\bar t_3)^{-1} & 0 \\
0      & 0 &  3t_3r_n
      \\
-|3t_3|^2r_nr_{n-1} &0 & 0
\end{array}
\right ) \end{matrix}
,\quad\quad\displaystyle
{\bf C}_n=\begin{matrix}
\left(
\begin{array}{ccc}
      0 &
0 & 0\\
       0& 0 &  0
      \\
0 & 0 & 3t_3
\end{array}
\right ). \end{matrix}
$$

\section{Semiclassical wave function}\la{Semi2}
In the semiclassical limit, $\hbar\to 0$, while
$t=\hbar N$, $t^{(\alpha )}=\hbar\nu_\alpha$, and
the potential $V(z)$ are kept fixed.
The quantum curve and evolution equations go to their
semiclassical counterparts.
Also, the semiclassical
wave functions are written through the objects of the
semiclassical theory, i.e., through differentials on
the semiclassical curve.

A technical difficulty is that wave
functions with close indices, say $\psi_n$ and
     $\psi_{n+1}$, are not necessarily close to each other
as $n \to \infty$.
Choosing a sequence of states
which has a semiclassical limit
is sensitive to the potential, and reflects the
configuration of semiclassical  droplets.
This problem does not arise in the
case of a single droplet in the algebraic case, where
states with close numbers are close. As we discussed above, this case
corresponds
to degenerate Riemann surfaces.
Hereafter,
we assume that states with close $n$'s are close, and proceed to the
semiclassical limit.
The result obtained for this case can be generalized to 
any smooth Riemann surface.

First, we rewrite the $L$ operator (\ref{M11}) in
  a more symmetric form
$L=\hat w^{-1/2} r \hat w^{1/2}+\sum \hat w^{-k/2} u^{(k)} \hat w^{-k/2}$,
by redefining $r$ and $u^{(k)}$. Then we search for a semiclassical  expansion
\be \nonumber
\psi(z)
\sim e^{\frac{1}{\hbar}{\mathcal A}_0 (z)+{\mathcal A}_1 (z) +\dots}
\ee
(in what follows, we suppress the index $n$).

In the first two leading orders, the  $L$-operator
(\ref{M11})  reads
$$L \longrightarrow z(w)+\frac{\hbar}{2}\{w
\p_wz(w),\p_t\}{\mathcal A}_1+\dots,$$
where $\log w=\p_t{\mathcal A}_0$ and $\{\,,\,\}$ means anticommutator.
The function
\be\la{681}
z(w)=r  w+\sum_{k\geq 0}u^{(k)} w^{-k}
\ee
     is obtained by  replacing the shift
operator  by its classical value $\hat w\to w$.

The recurrence relation
$L \psi(z)=z \psi(z)$ in two leading orders reads
\be\nonumber
z(w)=z,\quad
\p_t{\mathcal A}_1=-\frac{1}{2} \frac{\p_t \p_w z(w)}{\p_w z(w)},\quad
\log w=\p_t{\mathcal A}_0.
\ee
It gives
\be\la{70}
{\mathcal A}_1(z)=\frac{1}{2}\log w'(z),
\ee
where $w(z)$ is a multivalued  function inverse to (\ref{681}).

In the leading orders, (\ref{M}) reads
\be\la{711}
\hbar\p_z\psi(z)=\Big(-\frac{\bar z}{2}+\bar
z(w^{-1})+\frac{\hbar}{2}\{w\p_w\bar z(w^{-1}),\p_t\}+{\mathcal
A}_1+\dots\Big)\psi(z).
\ee
The function $\bar z(w^{-1})$ is the Schwarz reflection  of $z(w)$
with respect to the
contour whose exterior is mapped to the exterior of the unit disk
by the function $w(z)$. This map is discussed in Sec.~\ref{C}.
The contour is the boundary of a semiclassical droplet. We observe
that in the simplest semiclassical limit
the conformal map and its Schwarz reflection are
limits of the $L$ and $L^{\dag}$
operators of Sec.~\ref{1A}.
The complex curve
in the semiclassical limit
is given by $f(z,\bar z)=a(z)\prod (\bar z-\bar z(w^{-1}(z))$, where
the product is taken over
all possible branches of the multivalued function $w(z)$.

Introducing the Schwarz function $S(z)=\bar z(w^{-1}(z))$,
as in (\ref{ss}), we obtain
\be\nonumber
{\mathcal A}_0=-\frac{|z|^2}{2}+\Omega(z),
\ee
where $\Omega(z)=\int^z_{\xi_0} d\Omega$ is
the primitive function of the
generating differential
introduced in  Sec.~\ref{Diff}.
The point $\xi_0$ is such that $\log w(\xi_0)=0$. It is
a boundary point. The integral goes along the
physical sheet.

Summing up, we obtain the result:
\beq\label{sp1}
  \psi (z)\simeq (2\pi^3 \hbar )^{-1/4}\sqrt{w'(z)}\,
e^{\frac{1}{\hbar}(-\frac{1}{2}|z|^2 +\Omega (z))}.
\eeq

Formula (\ref{sp1}) was  reported in  \cite{ABWZ02}. A similar
    formula for the two-matrix model
was obtained in \cite{Eynard97}. Being
applied to the Hermite or Laguerre
polynomials
(see example in Sec.~\ref{classicalellipse}), the formula gives
   the asymptotical behaviour found by Tricomi in 1941 \cite{BE}.

The overall numerical factor can be fixed either by comparison with
asymptotics of biorthogonal
polynomials for the Ginibre ensemble, or
through normalization of the semiclassical wave function.
The latter is described below.

An extension of this formula
to the case of a general complex curve and
multiple droplets is
straightforward.
     The function $\Omega (z)$ should be again
understood as an integral of the generating
differential $d\Omega$. The integration path
starts at a point on the boundary
of some ``reference" droplet and ends at the point $z$.
The correction factor, $w'(z)$ in
(\ref{sp1}), is replaced by the
meromorphic Abelian differential of the third kind
$W^{(\infty,\bar\infty )}(z)$, introduced
  in Sec.~\ref{Diff}. As a
result, we obtain:
\be\la{75}
  \psi (z)\sqrt{dz}\sim \sqrt{W^{(\infty,\bar\infty )}(z)}\,
e^{-\frac{1}{\hbar}\big(\frac{|z|^2}{2}-\int^z_{\xi_0} S(z) dz\big)}.
\ee
By virtue of (\ref{18}), monodromies of this
wave function around droplets are $e^{2\pi i\nu_{\alpha}}$,
where $\nu_{\alpha}$ are filling factors.

The asymptotical representation (\ref{75}) fails
around cuts or double points. It is  valid only around boundaries of
physical droplets and
only if the distance between boundary of a physical droplet
and a cut or a double point
is much larger than $\sqrt\hbar$. In other words, it
is valid for smooth non-critical curves.
The semiclassical asymptotics of the wave functon in
the whole plane is far beyond the scope of this paper.

The following properties of ${\mathcal A}_0$ visualize the result.
\begin{itemize}
\item [(i)]The real part  of ${\mathcal A}_0$
is constant along boundary of any
droplet (see Sec.~\ref{Diff}), generally  different for different droplets
\be\nonumber
2{\mathcal R}e {\mathcal A}_0(z)=\phi_\alpha,
\quad  z\in\p D_\alpha.
\ee

\item[(ii)]Derivatives $\p_z$ and $\p_{\bar z}$
of ${\mathcal R}e\, {\mathcal A}_0$  vanish on the boundary:
\be \nonumber
\p_z {\mathcal R}e {\mathcal A}_0(z)=
\p_{\bar z} {\mathcal R}e {\mathcal A}_0(z)=
0, \quad z\in\p D.
\ee
This is the
condition for semiclassical extrema (\ref{s}).
\item[(iii)]
In the leading order, the change of
${\mathcal R}e \, {\mathcal A}_0$ under
small deviations along the normal direction away
from the boundary to both exterior
and interior does not depend on the point of the boundary:
\be \nonumber
\p_n^2 {\mathcal R}e\, {\mathcal A}_0(z)=-2.
\ee
\item [(iv)] The higher order expansion
away from the boundary depends of the curvature of the boundary $k(z), 
z\in\p D_\alpha$:
\beq\label{sp61}
{\mathcal R}e{\mathcal A}_0(z +\delta n )=\phi_\alpha - (\delta n)^2 +
\frac{1}{3}k(z)(\delta n)^3
-\frac{1}{4}k^2 (z) (\delta n)^4 +\ldots,
\eeq
where $\delta  n $ is a normal deviation from the boundary directed outward.
This expansion fails
at very curved parts of the contour, where $k(z)\sim \hbar^{-1/2}$.
\end{itemize}
     Altogether, these properties mean
that the wave function peaks at the
boundary of a physical droplet and
decays as Gaussian
   at the rate $\sqrt\hbar$ (see Fig.~\ref{semi}):
\beq\label{sp6}
|\psi (z\!+\!\delta n)|^2
\sim
e^{\frac{\phi_\alpha}{\hbar}}e^{-\frac{2}{\hbar}(\delta_n )^2}\,,
\;\;\;\;z\in \p D_\alpha.
\eeq
The amplitude of the wave function may be different on different
boundaries.

Since the
wave function is localized on the boundary, one
can integrate in the normal direction. In the case
of one droplet, we have
\be\la{78}
\int |\psi (z)|^2 d(\delta n)=
\frac{1}{2\pi}
\p_n\log |w(z)|=\frac{1}{2\pi}|w'(z)|,\quad z\in\p D
\ee
(here $\p_n$ is the normal derivative).
Further integration along the boundary is
consistent with the normalization
$\int |\psi_N (z)|^2 d^2z =1$.
This can be used to determine the constant in (\ref{sp1}).

According to the arguments of  Sec.~\ref{F1}, Eq. (\ref{78})
can be interpreted as a rate of growth.

Note that Eq. (\ref{70}) and the next to leading order
     of (\ref{711}) give the relation
\be\la{zz} w(\p_w z\p_t\bar z-\p_tz\p_w\bar z)=2,
\ee
where one differentiates at a fixed $w$. This relation
represents evolution of the growing domain
as a time-dependent conformal map.
The area of the domain grows linearly with time,  while
   its harmonic moments stay fixed. This type of classical
growth is equivalent
to the Laplacian growth.
Relation (\ref{zz}) has been known in the
literature for a long time \cite{Kochina}.
It is a classical limit of Eq. (\ref{string}).

\section{Appendix. Laplacian growth}\label{Appendix_A}
\begin{figure} \begin{center}
\includegraphics*[width=5cm]{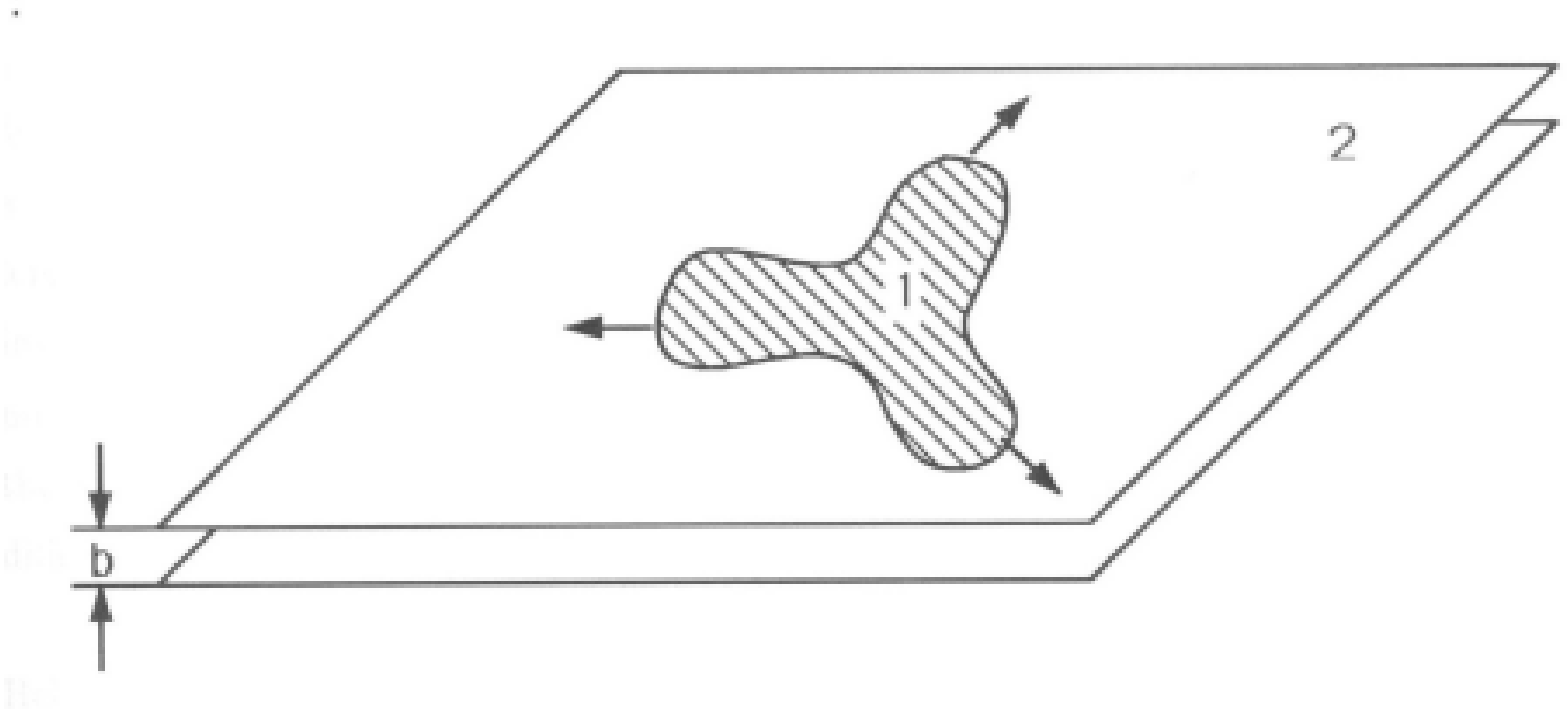}
\caption{\label{hs1}The Hele-Shaw setup.}
\includegraphics*[width=5cm]{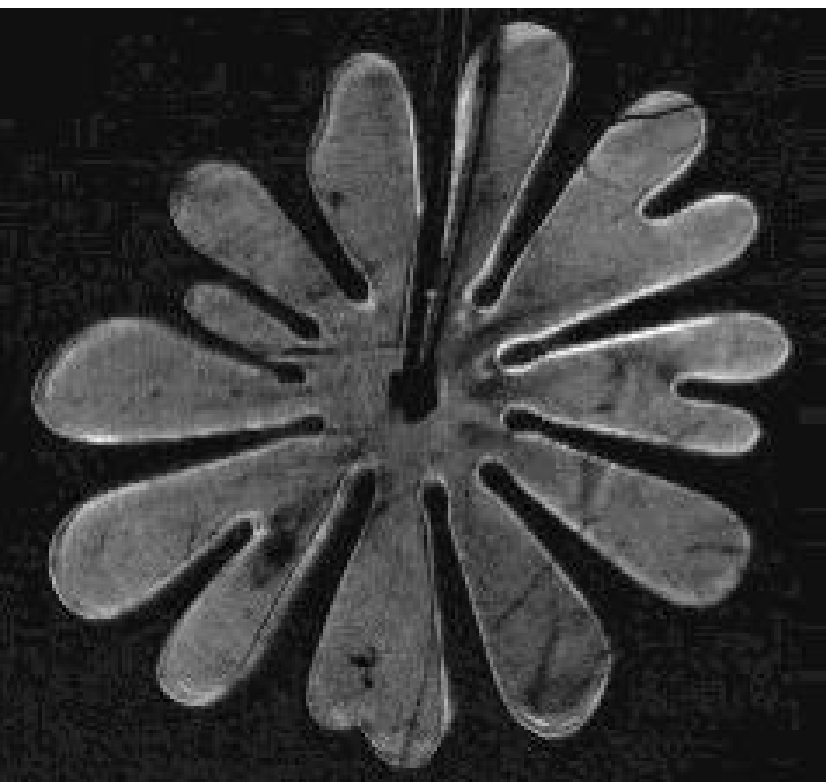}
\caption{\label{hs3}Laminar Hele-Shaw dynamics.}
\end{center} \end{figure}

Laplacian growth referres to growth of a planar  domain, whose boundary
propagates with a velocity
proportional to the gradient of a harmonic field. The
Hele-Shaw problem is a typical example (see \cite{review} for a
review). The experimental set-up consists (Fig.~\ref{hs1})
of two parallel and horizontal glass plates, separated by a small distance along the
vertical direction, called $the$ $gap$, $b$. The gap is small enough compared
to typical intermolecular distances in fluids, so that the system can be considered
two-dimensional.

Initially, the space between the plates is filled with fluid 2, of large viscosity
(tipical examples are silicon oil or liquid crystals). Through a small opening in
the middle of the upper glass plate, fluid 1, of much lower viscosity (such as air),
is inserted into the system, at a constant rate. Both fluids are incompressible and
immiscible, so fluid 2 must be evacuated from the system at the same rate as fluid
1 is being introduced. This is done somewhere at the edge of the glass plates.

Given that the fluids are incompressible and immiscible, the dynamics of the system
is reduced to that of the interface between them. Depending on the relevant
parameters of the experiment (the coefficients of viscosity, $\mu_{1,2}$ of the
fluids, the coefficient of surface tension $\sigma$, the gap $b$ and the pumping rate
$Q$), the interface dynamics will exhibit very different behaviors. At low pumping
rates $Q$, for given $\sigma$, the interface will remain smooth at all times, regardless
of how much fluid is being inserted (Fig.~\ref{hs3}). However, at high $Q$ (or
equivalently, at very low $\sigma$), the boundary will develop the so-called $fingering$
$instability$, where certain regions grow in the shape of fingers, whose tips split and
form secondary fingers, etc.  Real experiments (Fig.~\ref{sw})
show that in this regime,  the interface seems to evolve into a self-similar, fractal structure.

Under the assumption that the system is two dimensional, the equations
governing the dynamics of the interface lead to Darcy's Law:
the velocity field $\vec{v}_{i}$ and pressure $p_i$ in fluid $i$ are related as:
\be
\vec{v}_i = -\frac{b^2}{12\mu_i}\vec{\nabla}p_i,
\ee
and therefore the continuity equation accross the interface leads to
\be
\vec{v}\cdot \vec{n} = -\frac{b^2}{12\mu_1}\vec{n}\cdot\vec{\nabla}p_1
= -\frac{b^2}{12\mu_2}\vec{n}\cdot\vec{\nabla}p_2,
\ee
where $\vec{n}$ is the unit vector normal to the interface at a given point.

In each fluid, the pressure field is supposed to solve the Laplace equation,
\be
\Delta p_i = 0,
\ee
with appropriate boundary conditions. For instance, since fluid 2 is drained from the
system somewhere away from the origin, fluid 2 has a sink at infinity, which leads to
the condition
\be
p_2(z) \to -\log |z|, \quad z \to \infty.
\ee
The coefficient of surface tension enters the problem through the Laplace formula
relating the pressure jump accross the boundary and the local curvature $\kappa(z)$:
\be
p_2(z) -p_1(z) = \sigma \kappa(z).
\ee

The first simplification of the problem consists in setting the viscosity of fluid 1 to zero,
$\mu_1=0$. The continuity condition then implies $p_1 = $ constant, and we may therefore
choose $p_1 = 0$ (a redefinition of $p_2$).

The second important simplification is neglecting the effect of surface tension. This assumption
has important physical meaning, since the resulting problem should necessarily lead to turbulent
dynamics. Mathematically, the two assumptions lead to the following reformulation of the
problem:
\be
\Delta p = 0 \mbox{ on } D_{-}, \quad p = 0, \, \, V_n \sim \p_n p \, \, \, \mbox{ on } \Gamma,
\quad p(z) \stackrel{z \to \infty}{\longrightarrow} -\log|z| ,
\ee
where $V_n$ is the normal component of the velocity, $D_{-}$ is the domain occupied by fluid 2,
and $\Gamma$ is the boundary between fluids. Under these assumptions, in order to describe the
interface dynamics, we must solve an exterior Dirichlet problem for the pressure field.

\section*{Acknowledgments}

We are indebted to
A. Kapaev,
V. Ka\-za\-kov, I. Kri\-che\-ver, I. Kos\-tov,
A. Mar\-sha\-kov and M. Mi\-ne\-ev-\-Wein\-stein
for useful discussions, interest in the subject and help.
P.W. and R.T. were supported by the NSF MRSEC Program under
DMR-0213745, NSF DMR-0220198 and by the Humboldt foundation.
    A.Z. and P.W. acknowledge support
by the LDRD project 20020006ER ``Unstable
Fluid/Fluid Interfaces" at Los Alamos National Laboratory and M.
Mi\-ne\-ev-\-Wein\-stein for the hospitality
in Los Alamos. A.Z. was also supported in  part by  RFBR grant
03-02-17373 and by the grant for support of scientific schools
NSh-1999.2003.2. P.W.  is grateful to   K.B. Efetov  for the hospitality in
Ruhr-Universitaet Bochum  and to A.Cappelli for the hospitality in the
University of Florence,
where this work was completed. We are grateful to Harry Swinney
for permitting us to use Fig.~\ref{sw} from \cite{Sw}.


\begin{thebibliography}{10}
\bibitem{mwz} Mineev-Weinstein, M., Wiegmann, P.B., \& Zabrodin, A., 2000,
 Phys.\ Rev.\ Lett.\ \textbf{84}, 5106.

\bibitem{ABWZ02}  Agam, O., Bettelheim, E., Wiegmann, P.B.,
  \& Zabrodin, A., Viscous Fingering and a Shape of an
Electronic Droplet in a Quantum Hall Regime,  2002,     Phys. Rev. Lett.
{\bf {88}}, 236802.

\bibitem{mkwz}Krichever, I., Mineev-Weinstein, M., Wiegmann, P.B., \& 
Zabrodin, A., Laplacian growth and Whitham Equations of Soliton Theory,
2003, nlin.SI/0311005.

\bibitem{DV03}      Dijgraaf, R., \& Vafa, C.,  ${\mathcal{N}}=1$
Supersymmetry, Deconstruction and Bosonic Gauge Theory, 2003, hep-th/0302011,
A Perturbative Window into Non-Perturbative Physics, 2002, hep-th/0208048,
Matrix Models, Topological Strings and
Supersymmetric Gauge Theories, 2002, hep-th/0206255,
On Geometry and Matrix Models, 2002, hep-th/0207106.

\bibitem{BEH}
Bertola, M., Eynard, B.  \&  Harnad, J.,
Partition functions for Matrix Models and Isomonodromic
Tau Functions, 2002, nlin/0204054.

\bibitem{Kapaev} Kapaev, A., Riemann-Hilbert problem
for bi-orthogonal polynomials, 2003, J. Phys. A
{\bf 36}, 4629-4640.

\bibitem{gravity} Di Francesco, P., Ginsparg, P., \& Zinn-Justin, J.,
2-D Gravity and Random Matrices, 1995, Phys. Rept. 254, 1-133.

\bibitem{Its}Fokas, A.S., Its, A.R., \& Kitaev, A., The isomonodromy
 approach to the matrix models in 2-D gravity, 1992, Comm. Math. Phys. 147, 395-430.



\bibitem{KM03}      Kazakov, V.A., \& Marshakov, A.,
    Complex Curve of the Two Matrix Model and its Tau-function, 2003,
    J. Phys. A {\bf {36}}, 3107-3136.



\bibitem{Chekhov}Chekhov, L., \&  Mironov, A., Matrix
Models vs. Seiberg-Witten/Whitham theories, 2002,
hep-th/0209085.

\bibitem{David}David, F., Non-Perturbative Effects
in Matrix Models and Vacua of Two Dimensional
Gravity, 1993, Phys.Lett. B302, 403-410; hep-th/9212106; David, F., Bonnet, G., \&
Eynard, B., Breakdown of universality in multi-cut matrix models, 2000,
J.Phys. A33, 6739.


\bibitem{Zaboronsky}Chau, Ling-Lie, \& Zaboronsky, O., On the structure
of Normal Matrix Model, 1998, Commun.
Math. Phys. 196, 203-247, hep-th/9711091.
\bibitem{Mehta91}       Mehta, M.L., Random
Matrices, 1991,   Academic Press, New York.

\bibitem{WZ03}      Wiegmann, P.B., \& Zabrodin, A., Large
Scale Correlations in Normal and General Non-Hermitian Matrix
Ensembles, 2003, J. Phys. A, {\bf {36}}, 3411-3424.



\bibitem{Ginibre65}     Ginibre, J.,  Statistical
Ensembles of Complex Quaternion and Real Matrices, 1965, Journal of
Math. Phys. {\bf {6}} (3),  440.
  Girko, V.L., Elliptic Law, 1986, Theory of Probability and
Its Applications, 30, (4), 677-690.

\bibitem{Sw}Sharon, E.,
Moore, M.G., McCormick, W.D., \&  Swinney, H.L.,
Coarsening of Fractal  Viscous Fingering Patterns, 2003,
Phys. Rev. Lett., 91, 205504.


\bibitem{mesoscopic}Altshuler, B.L., \& Simons, B.D.,
Universalities: from Anderson Localization to Quantum Chaos,
in: Mesoscopic Quantum Physics, 1994, Les Houches 1994, eds.
Akkermans, E.,  Montambaux, G.,  Pichard, J-L., \&  Zinn-Justin, J.



\bibitem{Aratyn}Aratyn, H., Integrable Lax Hierarchies, their
Symmetry Reductions
and Multi-Matrix Models, 1995, hep-th/9503211.


\bibitem{BEHdual}
Bertola, M., Eynard B., \&  Harnad, J.,
Duality of spectral curves arising in two-matrix models, 2001,
nlin/0112006.

\bibitem{Eynard03}  Bertola, M., Eynard, B., \& Harnad, J.,
    Differential systems for biorthogonal polynomials appearing in 2-matrix
models and the associated
Riemann-Hilbert problem, 2002, nlin/0208002.

\bibitem{Krichev-red} Krichever, I.M., Funkt. Analiz i
ego Pril., 1995, {\bf 29/2}, 1-8.

\bibitem{Wasow}
Wasow, W., Asymptotic Expansions for Ordinary
Differential Equations, 1965, John Wiley \& Sons, New York.

\bibitem{curve}Dubrovin, B.A., Krichever, I.M., \& Novikov, S.P.,
Topological and
algebraic geometry methods in contemporary mathematical physics II, 1982,
  Soviet Scient. Reviews, Math. Phys. Reviews 3 1-150;
Krichever, I.M., Integration of nonlinear equations by the method
of algebraic geometry, 1977, Funct. Anal. Appl., {\bf 11}, 12-26.


\bibitem{Alhfors50}     Ahlfors, L.V., Complex
Analysis, an Introduction to the Theory of Analytic Functions of One
Complex Variable, 1953, McGraw-Hill, New York.


\bibitem{C}
Cohn, H., Conformal Mapping on Riemann Surfaces, 1967, Dover, New York.


\bibitem{SS}  Schiffer, M., \& Spencer, D.C.,
   Functionals of finite Riemann surfaces, 1954,
Princeton University Press.


\bibitem{Miwa}Jimbo, M., Miwa, T., \& Ueno, K.,
Monodromy preserving deformations of
linear ordinary differential equations with rational coefficients I, 1981, Physica D, {\bf 2}, 306-52; Jimbo, M., \& Miwa, T.,
Monodromy preserving deformation of linear
ordinary differential equations with rational coefficients II, 1981, Physica D,
{\bf 2}, 407-48; Flaschka, H., \& Newell, A.C.,
Monodromy- and spectrum-preserving
deformations, 1980, Commun. Math. Phys., {\bf 76}, 65-116.

\bibitem{Takasaki} Takasaki, K.,
Dual Isomonodromic Problems and Whitham Equations, 1998, Lett. Math. Phys.
{\bf 43} 123-135.


\bibitem{Ahar-Shap}
Aharonov, D., \& Shapiro, H., 1976,
J. Anal. Math., {\bf 30},  39-73;\\
Shapiro, H., The Schwarz function
and its generalization to higher dimensions, 1992,
University of Arkansas Lecture Notes in the
Mathematical Sciences, Volume 9, Summers W.H.,
  Editor, John Wiley \& Sons.

\bibitem{Kostov}Kostov, I.K., Conformal field theory techniques in
random matrix models,  1999, arXiv:hep-th/9907060.


\bibitem{review} Bensimon, D.,  Kadanoff, L.P.,
Liang, S., Shraiman, B.I., \& Tang, C., 1986,
Rev.\ Mod.\ Phys.\, \textbf{58}, 977.

\bibitem{kkmwz}Kostov, I.K., Krichever, I., Mineev-Weinstein, M.,
Wiegmann, P.B., \& Zabrodin, A., $\tau$-Function for Analytic
Curves, in:
Random matrices and their applications, 2000, MSRI publications, vol. 40, 285,
Cambridge  University Press.

\bibitem{Eynard97}      Eynard, B., Eigenvalue
Distribution of Large Random Matrices, from One Matrix to Several
Coupled Matrices, 1997, cond-mat/9707005.


\bibitem{BE} Bateman, H., \&  Erdelyi, A., Higher transcendal
functions, 1953,
v. 2, McGraw-Hill.


\bibitem{Morozov}Marshakov, A.,  Mironov, A., \&  Morozov, A.,
  Generalized matrix models as conformal
field theories: Discrete case, 1991, Phys. Lett. B, {\bf 265}, 99;  Kharchev, S.,
  Marshakov, A.,  Mironov, A.,  Morozov, A., \& Pakuliak, S.,
Conformal matrix models
as an alternative to conventional multimatrix models, 1993, Nucl. Phys. B
{\bf 404}, 717, [arXiv:hep-th/9208044].



\bibitem{Toda}
Ueno, K., \& Takasaki, K., 1984,
Adv. Stud. Pure Math., {\bf 4}, 1.



\bibitem{Whitham}Whitham, J.B.,
Linear and nonlinear waves, Wiley-Interscience, 1974,
New York;
Flaschka, H., Forest, M.G., \& McLaughlin, D.W.,
  Multiphase averaging and the inverse spectral solution of 
the Korteweg-de Vries equation, 1980,
Comm. Pure Appl. Math., {\bf 33}, 739-84;
Krichever, I.M.,
Method of averaging for two-dimensional ``integrable"
equations, 1988, Funct. Anal. Appl., {\bf 22}, 200-213.

\bibitem{Kochina}
Galin, L.A., 1945, Dokl. Akad. Nauk SSSR,
    {\bf 47},  250-253;\\
Polubarinova-Kochina, P.Ya., 1945, Dokl. Akad. Nauk SSSR,
    {\bf 47},  254-257; \\
Kufarev, P.P., 1947, Dokl. Akad. Nauk SSSR
{\bf 57},  335-348.


\bibitem{Akemann02}Akemann, G., The Solution of a Chiral Random
Matrix Model with Complex Eigenvalues, 2002, J. Phys. {\bf A36}, 3363.


\bibitem{1}Wiegmann, P.B., \& Zabrodin, A.,
Conformal maps and dispersionless integrable hierarchies, 2000,
Commun.Math.Phys. {\bf 213},  523-538;
Marshakov, A., Wiegmann, P.B., \& Zabrodin, A.,
Integrable Structure of the Dirichlet Boundary Problem in Two Dimensions, 2002,
Commun. Math. Phys. {\bf 227}, 131-153.
\bibitem{mkz} Krichever, I., Marshakov, A., \& Zabrodin, A.,
Integrable Structure of the Dirichlet Boundary Problem in
Multiply-Connected Domains, 2003, hep-th/0309010.

\end{thebibliography}
\end{document}